\documentclass[12pt]{article}
\usepackage{a4}
\usepackage{latexsym} % Gets \Box etc
\usepackage{amssymb}  % \gtrsim, \geqslant, etc etc: 
\usepackage{amsmath} 
\usepackage{graphicx}
\usepackage{epsfig,wrapfig}
\usepackage[ usenames]{color}
\usepackage[rflt]{floatflt} 
\begin{document}

%\voffset-3cm
%%%%%%%%%%%%%%%%%%%%%%%%%%%%%%%%%%%%%%%%%%%%%%%%%%%%%%%%%%%%%%%%%%%%%%%%
% several bibliographies in one text file
% the series of new commands, tested on LaTeX2e and on Latex2.09
% via the compatibility mode.
% version redefining thebibliography and cite.
% J. Bijnens 18 Sep. 1998
%%%%%%%%%%%%%%%%%%%%%%%%%%%%%%%%%%%%%%%%%%%%%%%%%%%%%%%%%%%%%%%%%%%%%%%%
\newcommand{\talk}[3]
{\noindent{#1}\\ \mbox{}\ \ \ {\it #2} \dotfill {\pageref{#3}}\\[1.8mm]}
\newcommand{\stalk}[3]
{{#1} & {\it #2} & {\pageref{#3}}\\}
\newcommand{\snotalk}[3]
{{#1} & {\it #2} & {{#3}n.r.}\\}
\newcommand{\notalk}[3]
{\noindent{#1}\\ \mbox{}\ \ \ {\it #2} \hfill {{#3}n.r.}\\[-4mm]}
\newcounter{zyxabstract}     %  step by one when going to new abstract
\newcounter{zyxrefers}        %  counts the references in every bibliography

\newcommand{\newabstract}
{\newpage\stepcounter{zyxabstract}\setcounter{equation}{0}
\setcounter{footnote}{0}\setcounter{figure}{0}\setcounter{table}{0}}

\newcommand{\rlabel}[1]{\label{zyx\arabic{zyxabstract}#1}}
\newcommand{\rref}[1]{\ref{zyx\arabic{zyxabstract}#1}}

\renewenvironment{thebibliography}[1] % as standard,the width of largest number
{\section*{References}\setcounter{zyxrefers}{0}
\begin{list}{ [\arabic{zyxrefers}]}{\usecounter{zyxrefers}}}
{\end{list}}
% for those who really need every possible mm:
\newenvironment{thebibliographynotitle}[1] % as 
%                              standard,the width of largest number
{\setcounter{zyxrefers}{0}
\begin{list}{ [\arabic{zyxrefers}]}
{\usecounter{zyxrefers}\setlength{\itemsep}{-2mm}}}
{\end{list}}

\renewcommand{\bibitem}[1]{\item\rlabel{y#1}}% extra y to avoid duplication
\renewcommand{\cite}[1]{[\rref{y#1}]}      %of labels in text and bibliography
\newcommand{\citetwo}[2]{[\rref{y#1},\rref{y#2}]}
\newcommand{\citethree}[3]{[\rref{y#1},\rref{y#2},\rref{y#3}]}
\newcommand{\citefour}[4]{[\rref{y#1},\rref{y#2},\rref{y#3},\rref{y#4}]}
\newcommand{\citefive}[5]
{[\rref{y#1},\rref{y#2},\rref{y#3},\rref{y#4},\rref{y#5}]}
\newcommand{\citesix}[6]
{[\rref{y#1},\rref{y#2},\rref{y#3},\rref{y#4},\rref{y#5},\rref{y#6}]}
%%%%%%%%%%%%%%%%%%%%%%%%%%%%%%%%%%%%%%%%%%%%%%%%%%%%%%%%%%%%%%%%%%%%%%%%%%%%

\begin{titlepage}

\begin{flushleft}
\includegraphics[height=2.5cm]{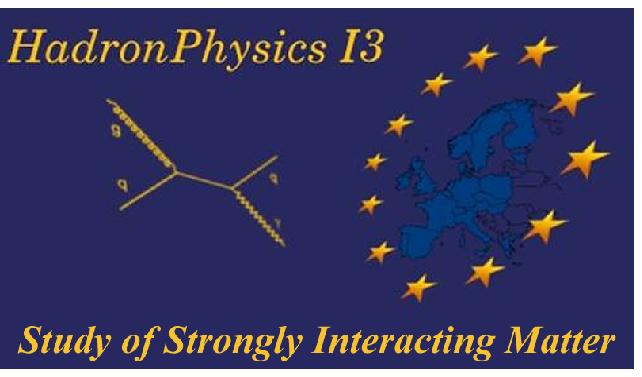} 
\end{flushleft}

\vspace*{-1.75cm}
\begin{flushright}
\small{\tt HISKP-TH-06/28}
\end{flushright}

\vspace*{1.cm}

\begin{center}
{\large \bf Workshop on exotic hadronic atoms, deeply bound kaonic 
nuclear states 
and antihydrogen:
present results, future challenges}\\[0.5cm]
{ECT*, Strada delle Tabarelle 286, I-38050, Villazzano (Trento), Italy}\\
{June 19-24, 2006}
\\[1cm]
{\em edited by}\\[1cm]
{\bf C. Curceanu (Petrascu)$^1$, A. Rusetsky$^{2,3}$, E. Widmann$^4$}\\[0.3cm]
{\em $^1$LNF - INFN, Via E. Fermi 40, 00044 Frascati, (Roma), Italy}\\
{\em $^2$Universit\"{a}t Bonn, Helmholtz-Institut f\"{u}r
Strahlen- und Kernphysik (Theorie)\\ Nu{\ss}allee 14-16, D-53115 Bonn, Germany}\\
{\em $^3$On leave of absence from: 
HEPI, Tbilisi State University\\ University st. 9, 380086 Tbilisi, Georgia}\\
{\em $^4$Stefan Meyer Institut f\"{u}r subatomare 
Physik, Boltzmanngasse 3, A-1090, Vienna, Austria}

\vspace*{.2cm}

\begin{figure}[h]
\begin{center}
\includegraphics[height=5.cm]{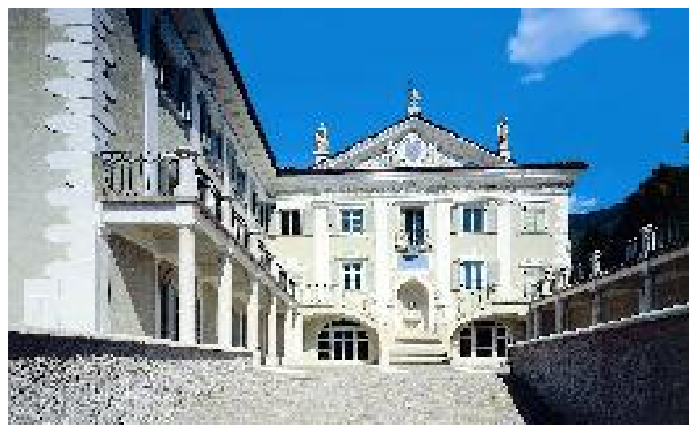} 
\end{center}
\end{figure}

{\large ABSTRACT}
\end{center}
These are the miniproceedings of the workshop {\it ``Exotic hadronic atoms, 
deeply bound kaonic nuclear states and antihydrogen:
present results, future challenges,''} which was held at
the European Centre for Theoretical Nuclear Physics and Related
Studies (ECT*), Trento (Italy), June 19-24, 2006.
The document includes a short presentation of the topics,
 the list of participants,  and
a short contribution from each speaker.

\end{titlepage}

\setcounter{page}{2}

\newabstract

\section{Introduction}

%\vspace{1cm}

The workshop ``Exotic hadronic atoms, 
deeply bound kaonic nuclear states and antihydrogen:
present results, future challenges'' was held on June 19-24 
at the European Centre
for Theoretical Nuclear Physics and Related Topics (ECT*), Trento (Italy).
The workshop has been largely inspired by latest theoretical 
and experimental
progress, achieved in the investigation of the exotic hadronic bound systems,
which offer a unique way for studies of fundamental interactions and 
symmetries in Nature.
The topics of the workshop included:

\begin{sloppypar}

\begin{itemize}

\item 
Hadronic atoms: status of the theory and of experimental results 
(including DEAR/SIDDHARTA at DAFNE, DIRAC at CERN, Pionic atoms at PSI) 

\item
Meson-nucleon and meson-nucleus interactions in effective theories, 
unitarization models of Chiral Perturbation Theory.

\item
$K\to 3\pi$ decay: theory and experimental results (including NA48 at CERN)

\item
Deeply bound kaonic-nuclear states: status of the theory and 
experimental results (including E471 and E549/570 at KEK, FINUDA at 
DAFNE, FOPI at GSI) 

\item
Antihydrogen physics: production mechanisms, precision spectroscopy for 
testing CPT and QED, gravitation of antimatter 

\item
Next generation experiments at  DAFNE2, J-PARC, FAIR 

\end{itemize}

\end{sloppypar}

\noindent
Around 50 physicists took part in the workshop, which was held in a unique
atmosphere of intense discussions and learning. In total 45 talks were 
presented.  At the end of the workshop, an informal discussion on deeply
bound $\bar K$-nuclear states took place (convener: W. Weise).

\vspace{0.5cm}
\noindent {\em Acknowledgments.}

\bigskip

\noindent
We wish to thank all participants for traveling to Trento and for making
an exciting and very lively meeting. We want to thank our secretary
Ines Campo for the excellent performance, and the whole staff of the ECT*
for their help. 

\bigskip

\noindent
The Workshop was financially supported by ECT* and by the FP6 EC
contract RII3-CT-2004-506078 ``Study of strongly interacting matter" (``I3
HadronPhysics"). 

\vspace{0.5cm}
For the full program and the complete list of speakers
see:

http://www.itkp.uni-bonn.de/$\sim$rusetsky/TRENTO06/trento06.html

\bigskip\bigskip

\noindent Frascati/Bonn/Vienna, October 13, 2006

\bigskip\bigskip\bigskip

\hspace*{5.4cm} 
\begin{tabular}{l l}
Catalina Curceanu & (petrascu@lnf.infn.it)\\
Akaki Rusetsky & (rusetsky@itkp.uni-bonn.de)\\
Eberhard Widmann & (eberhard.widmann@oeaw.ac.at)
\end{tabular}

\newpage

\begin{sloppypar}

\section{List of Participants}
{\sf
\begin{tabbing}
Noxx\=A very long namexxxxx\=a very long institutexxxxxx\=email\kill
1.\> Petros Aslanyan\> (Yerevan) \>aslanyan@sunhe.jinr.ru \\
2.\> Yoshinori Akaishi\> (RIKEN) \> yoshinori.akaishi@kek.jp\\
3.\> Vadim Baru \>(ITEP) \> baru@itep.ru\\
4.\> George Beer \>(TRIUMF, Uvic) \> gbeer@uvic.ca\\
5.\> Bugra Borasoy \>(Bonn) \> borasoy@itkp.uni-bonn.de\\
6.\>Paul B\"uhler \>(SMI Vienna) \> paul.buehler@oeaw.ac.at\\
7.\> Gianmaria Collazuol \>(Pisa) \> gianmaria.collazuol@pi.infn.it \\
8.\>Catalina Curceanu \>(LNF-INFN) \> catalina@lnf.infn.it\\
9.\> Evgeny Epelbaum \>(FZ J\"ulich) \> epelbaum@itkp.uni-bonn.de\\
10.\> Torleif Ericson\>(CERN) \> torleif.ericson@cern.ch \\
11.\>Eli Friedman \>(Jerusalem) \> elifried@vms.huji.ac.il\\
12.\> Avraham Gal \>(Jerusalem) \>AVRAGAL@vms.HUJI.AC.IL \\
13.\> Juerg Gasser \>(Bern) \> gasser@itp.unibe.ch\\
14.\> Detlev Gotta \>(FZ J\"ulich) \> d.gotta@fz-juelich.de\\
15.\> Carlo Guaraldo \>(LNF-INFN) \> carlo.guaraldo@lnf.infn.it\\
16.\> Ryugo Hayano \>(Tokyo) \> ryugo.hayano@cern.ch\\
17.\>Norbert Herrmann \>(Heidelberg) \> herrmann@physi.uni-heidelberg.de\\
18.\> Gino Isidori \>(LNF-INFN) \> gino.isidori@lnf.infn.it\\
19.\> Andrey Ivanov \>(Vienna) \> ivanov@kph.tuwien.ac.at\\
20.\>Masahiko Iwasaki \>(RIKEN) \> masa@riken.jp\\
21.\> Bertalan Juhasz \>(SMI Vienna) \> Bertalan.Juhasz@cern.ch\\
22.\> Kenta Itahashi \>(RIKEN) \> itahashi@riken.jp\\
23.\>Paul Kienle \>(SMI Vienna) \> Paul.Kienle@ph.tum.de\\
24.\>Tadafumi Kishimoto \>(Osaka) \>  kisimoto@phys.sci.osaka-u.ac.jp\\
25.\>Alexander Kudryavtsev \>(ITEP) \> kudryavt@itep.ru\\
26.\>Naofumi Kuroda \>(RIKEN) \> \\
27.\> Ralf Lehnert \>(MIT) \> rlehnert@lns.mit.edu\\
28.\> Richard Lemmer \>(Witwatersrand) \> LemmerR@physics.wits.ac.za\\
29.\> Johann Marton \>(SMI, Vienna) \> johann.marton@oeaw.ac.at\\
30.\> Nikolaos Mavromatos \>(London) \> nikolaos.mavromatos@kcl.ac.uk\\
31.\>Natalia Troitskaya \>(St. Petersburg) \> natroitskaya@yandex.ru\\
32.\> Robin Ni\ss ler \>(Bonn) \> rnissler@itkp.uni-bonn.de\\
33.\> Hiroaki Ohnishi \>(RIKEN) \> h-ohnishi@riken.jp\\
34.\> Eulogio Oset \>(Valencia) \> oset@ific.uv.es\\
35.\>Jose A. Oller \>(Murcia) \>oller@um.es \\
36.\>Haruhiko Outa \>(RIKEN) \> outa@riken.jp\\
37.\>Joaquim  Prades \>(Granada) \> prades@ugr.es\\
38.\> Akaki Rusetsky \>(Bonn) \> rusetsky@itkp.uni-bonn.de\\
39.\> Udit Raha \>(Bonn) \> udit@itkp.uni-bonn.de\\
40.\> Leopold Simons \>(PSI) \> leopold.simons@psi.ch\\
41.\>Ludwig Tauscher \>(Basel) \> \\
42.\> Dirk Trautmann \>(Basel) \> dirk.trautmann@unibas.ch\\
43.\> Volodymyr Magas \>(Barcelona) \> vladimir@ecm.ub.es\\
44.\> Wolfram Weise \>(TUM Munchen) \> wolfram.weise@physik.tu-muenchen.de\\
45.\> Eberhard Widmann \>(SMI Vienna) \> eberhard.widmann@oeaw.ac.at\\
46.\> Slawomir Wycech \>(Warsaw) \> Slawomir.Wycech@fuw.edu.pl\\
47.\>Toshimitsu Yamazaki \>(Tokyo) \> yamazaki@nucl.phys.s.u-tokyo.ac.jp\\
48.\> Valeriy Yazkov \>(Moscow State) \> yazkov@nusun.jinr.ru\\
49.\>Johann Zmeskal \>(SMI, Vienna) \>Johann.Zmeskal@oeaw.ac.at \\
\end{tabbing}
}
\end{sloppypar}

\newpage

\section{Contributions}

\vskip.5cm

\noindent\mbox{}\hfill{\bf Page}

{\sc Exotic Atoms}

\noindent\hrulefill

\talk{{\bf L. M. Simons}}{Pionic Hydrogen}{abs:Simons}
\talk{{\bf D. Gotta}}{Pionic Deuterium}{abs:Gotta_deuterium}
\talk{{\bf E. Epelbaum}}{Chiral forces and few-nucleon systems}{abs:Epelbaum}
\talk{{\bf A. Rusetsky et al}}{The theory of pionic deuterium: status and perspectives}{abs:Rusetsky}
\talk{{\bf V. Baru et al}}{ChPT for $NN\to NN\pi $ and absorption correction to
$a_{\pi d}$}{abs:Baru}
\talk{{\bf K. Itahashi et al}}{Future programs for the precision spectroscopy
  of pionic atoms in the \\ \hspace*{.3cm} nuclear reactions}{abs:Itahashi}
\talk{{\bf B. Borasoy}}{Low-energy $\bar{K} N$ interactions}{abs:Borasoy}
\talk{{\bf J. A. Oller}}{About the Strangeness $-1$ S-wave Meson-Baryon Scattering}{abs:Oller}
\talk{{\bf R. Ni\ss ler}}{Chiral unitary approach to $K^- p$ scattering}{abs:Nissler}
\talk{{\bf J. Marton}}{Kaonic Hydrogen Experiments}{abs:Marton}
\talk{{\bf V. Yazkov}}{Measurement of the $\pi^+\pi^-$ atom lifetime at DIRAC}{abs:Yazkov}
\talk{{\bf U. Raha et al}}{$\bar{K}N$ Scattering Lengths From Kaonic Deuterium}{abs:Raha}
\talk{{\bf D. Trautmann et al}}{How accurate are the pionium breakup calculations?}{abs:Trautmann}
\talk{{\bf A. Kudryavtsev}}{ Analytic theory for hadronic atoms beyond the Deser approximation}{abs:Kudryavtsev}
\talk{{\bf T.E.O. Ericson et al}}{Corrections to scattering lengths from hadronic atoms}{abs:Ericson}
\talk{{\bf E. Friedman}}{Kaonic atoms as a starting point for antikaon nuclear physics}{abs:Friedman}

\newpage

{\sc Antiprotonic atoms}

\noindent\hrulefill

\talk{{\bf R.S. Hayano}}{Determination of the Antiproton-to-Electron Mass Ratio 
by Precision Laser\\ \hspace*{.3cm} Spectroscopy of $\bar{p}$He$^+$}{abs:Hayano_helium}
\talk{{\bf D. Gotta}}{Light Antiprotonic Atoms}{abs:Gotta_antiproton}
\talk{{\bf N.E. Mavromatos}}{Looking for ``smoking gun'' signatures of CPT Violation}{abs:Mavromatos}
\talk{{\bf R. Lehnert}}{Lorentz and CPT tests with antimatter}{abs:Lehnert}
\talk{{\bf R.S.Hayano}}{Antihydrogen}{abs:Hayano_antiproton}
\talk{{\bf B. Juhasz et al}}{Measurement of the ground-state hyperfine structure of antihydrogen}{abs:Juhasz}
\talk{{\bf N. Kuroda et al}}{MUSASHI -- An ultra-slow antiproton beam source -- 
Ultra-slow antiproton\\ \hspace*{.3cm}  beam source and antiprotonic atom formation}{abs:Kuroda}

\vspace*{2.cm}

{\sc $K$-clusters}

\noindent\hrulefill

\talk{{\bf S. Wycech}}{On the structure of KNN, KNNN states}{abs:Wycech}
\talk{{\bf A. Gal}}{Dynamical calculations of $K^-$ nuclear bound states}{abs:Gal}
\talk{{\bf E. Oset et al}}{Chiral dynamics of $\bar{K}$-nucleus interaction: critical review
 of deeply bound states}{abs:Oset}
\talk{{\bf Y. Akaishi et al}}{Deeply bound kaonic nuclear states in reply to recent criticisms}{abs:Akaishi}
\talk{{\bf A. Ivanov et al}}{Phenomenological quantum field theoretic
    model of Kaonic Nuclear Clusters\\ \hspace*{.3cm} 
    $K^-pp$, $K^-pnn$ and so on}{abs:Ivanov}
\talk{{\bf V. Magas et al}}{Simulation of the K- nuclear absorption at FINUDA}{abs:Magas}
\talk{{\bf P. Kienle}}{Probing the Structure of Nuclei Bound by Antikaons}{abs:Kienle}
\talk{{\bf W. Weise}}{Conditions for antikaon-nuclear bound states}{abs:Weise}
\talk{{\bf T. Yamazaki et al}}{Present status of the experimental investigation of deeply bound
 kaonic states}{abs:Yamazaki_deeplybound}
\talk{{\bf T. Kishimoto}}{Study of kaonic nuclei by in-flight $(K^-, N)$ reactions}{abs:Kishimoto}
\talk{{\bf N. Herrmann}}{Search for deeply bound kaonic states with FOPI at GSI}{abs:Herrmann}
\talk{{\bf P. B\"uhler et al}}{Search for K$^-$pp clusters in p+d-reaction with FOPI}{abs:Buehler}
\talk{{\bf P.Zh. Aslanyan}}{$\Lambda p$ spectrum analysis at 10 GeV/c in p+C interactions}{abs:Aslanyan}
\talk{{\bf T. Yamazaki et al}}{Enhanced formation of $K^-pp$ clusters by short-range $pp$
 collisions}{abs:Yamazaki_clusters}
\talk{{\bf H. Ohnishi}}{A search for deeply-bound kaonic nuclear states 
by in-flight $^3$He(K$^-$,n) reaction\\ \hspace*{.3cm}  at J-PARC}{abs:Ohnishi}
\talk{{\bf J. Zmeskal}}{AMADEUS  AT  DA$\Phi$NE}{abs:Zmeskal}
\talk{{\bf W. Weise (convener)}}{Discussion panel on deeply bound $\bar{K}$-nuclear states}{abs:Weise_summary}

\vspace*{2.cm}

{\sc $K\to 3\pi$ decays}

\noindent\hrulefill

\talk{{\bf J. Prades et al}}{FSI in $K\to 3\pi$  
and Cabibbo's Proposal to Measure $a_0-a_2$}{abs:Prades}
\talk{{\bf G. Collazuol}}{Pion scattering lengths from the NA48/2 experiment at CERN}{abs:Collazuol}
\talk{{\bf G. Isidori et al}}{On the cusp effects in $K \to 3\pi$ decays}{abs:Isidori}
\talk{{\bf J. Gasser et al}}{Non relativistic QFT and $K\rightarrow 3\pi$ decays}{abs:Gasser}

%%%%%%%%%%%%%%%%%%%%%%%%%%%%%%%%%%%%%%%%%%%%%%%%%%%%%%%%%%%%%%%%%%%%%%%%%

\newabstract %1 Simons
\label{abs:Simons}

\begin{center}
{\large\bf Pionic Hydrogen}\\[0.5cm]
 L.~M.~Simons\\[0.3cm]
 Paul Scherrer Institut, CH-5232 PSI/Villigen, Switzerland\\[0.3cm]
{\it for the PIONIC HYDROGEN collaboration}
 \end{center}
The new pionic hydrogen experiment at PSI \cite{R98.01} aims at an improvement in the 
determination of the strong interaction ground state shift and width of the pionic 
hydrogen atom using high precision X-ray crystal spectroscopy. The final goal 
is the extraction of isospin separated scattering lengths with accuracies at 
the percent level. Compared to previous efforts the energy resolution and 
statistics have been improved and the background is much reduced. The spectrometer 
response function has been determined precisely using a novel method\,\cite{Ana02}. 

The inherent difficulties of the exotic atom's method  result from the fact 
that the pionic hydrogen atom must be formed at higher gas pressures. For the 
extraction of a strong interaction shift an extrapolation method 
to vacuum conditions proved to be  successful and  the measured line shift 
\begin{equation}
\epsilon_{1s}=+7.120\pm 0.008\,(stat.)\pm 0.006\,(sys.)~eV 
\end{equation}
could be attributed exclusively to the strong interaction\,\cite{Hen03}.
The measured line shape of a pionic hydrogen K transition does not permit to
extract the strong interaction width $\Gamma_{1s}$ directly as it is 
Doppler broadened by various de-excitation steps caused by  by  $n\rightarrow n'$ 
Coulomb transitions.
Based on the precise knowledge of the spectrometer's resolution function it was  tried
to identify various contributions to 
the line shape from Coulomb de--excitation.

With this procedure the analysis of the three transitions
 $\pi H(2p-1s)$, $\pi H(3p-1s)$ and $\pi H(4p-1s)$  at different pressures 
could be combined to yield a preliminary result for the strong interaction width to be  
\begin{equation}
\Gamma_{1s}= 823\pm 19~meV.
\end{equation}
 
The efforts to improve the accuracy of the scattering lengths as well face the problem that
the linear combination $a^{+}+a^{-}$ to be determined from $\it\epsilon_{1s}$ 
suffers from the poor knowledge of $\delta_{\it\epsilon}$ \cite{Gasser_Ericson}. 
The correction $\delta_{\it\Gamma}$ for $\it\Gamma_{1s}$
seems to be much better under control \cite{Zem03}.This allows 
 to quote a preliminary value for  the isovector scattering length
to be 
\begin{equation}
a^{-} = (86.44\,^{\,+0.10\,}_{\,-1.02})\cdot 10^{-3}~[m^{-1}_{\pi}].
\end{equation}

\newabstract %2 Gotta_deuterium
\label{abs:Gotta_deuterium}
 \begin{center}
 {\large\bf Pionic Deuterium}\\[0.3cm]
 D. Gotta\\[0.3cm]
 Institut f\"ur Kernphysik, Forschungszentrum J\"ulich, D-52425 J\"ulich, Germany\\[0.3cm]
{\it for the PIONIC HYDROGEN collaboration}
 \end{center}
 
Precise measurements of shift $\epsilon_{1s}$ and width $\Gamma_{1s}$ of the 
pionic hydrogen ground state have been performed\,\cite{Simons} 
in order to extract the isoscalar and isovector scattering lengths $a^{+}$ and 
$a^{-}$ within the framework HB$\chi$PT calculations\,\cite{ChiPT}. Whereas 
$\Gamma_{1s}\propto (a_{\pi^{-}p}\rightarrow a_{\pi^{0}n})^{2}\propto (a^{-})^{2}$
yields directly $a^{-}$, the shift is due to elastic scattering with 
$\epsilon_{1s}\propto  a_{\pi^{-}p}\rightarrow a_{\pi^{-}p}\propto a^{+}+a^{-}$.
The determination of $a^{+}\ll a^{-}$ from this linear combination suffers in particular
from the large uncertainty of the low-energy constant $f_{1}$.

In $\pi D$ to leading order, $\epsilon_{1s}\propto a^{+}$. Consequently, higher orders 
(multiple scattering, deuteron structure, absorption, etc.) contribute significantly, 
but as shown by theoretical calculations are well under control\,\cite{piDtheo}. Hence, 
a precise measurement of $\epsilon_{1s}$ in $\pi D$ being an independent access to 
$a^{+}$ imposes constraints on $a^{+}$ and $a^{-}$ as obtained from $\pi H$. Even more, 
limits can be derived for $f_{1}$. It is worth mentioning that due to the leading order 
cancellations, isospin--breaking effects amount to $\approx 40\%$ for $\epsilon_{1s}$ 
in $\pi D$\,\cite{Mei05}, which is an outstanding occurance in pion--nuclear interaction 
involving charged pions.

For $\pi D$, $\Gamma_{1s}$ is directly related to threshold pion production 
in the reaction $pp\rightarrow \pi^{+}d$ by detailed balance and charge independence. 
The threshold parameter $\alpha $ representing s--wave pion production is proportional to the 
imaginary part of the $\pi d$ scattering length being $\propto \Gamma_{1s}$. Calculations 
of $\alpha $ within HB$\chi$PT are continuously improved and require experimental data 
at least at the expected final level of accuracy\,\cite{Len06}.

The forthcoming experiment, using the $\pi D(3p-1s)$ transition, aims at an 
improvement for $\epsilon_{1s}$ and $\Gamma_{1s}$ from $3\%\rightarrow 0.5\%$ 
and $12\%\rightarrow 4\%$, respectively\,\citetwo{ExppiD}{PSIpiD}. Possible molecular effects 
will be identified by studying the pressure dependence like in the $\pi H$ experiment.

\newabstract %3 Epelbaum
\label{abs:Epelbaum}
 \begin{center}
 {\large\bf Chiral forces and few-nucleon systems}\\[0.3cm]
 E. Epelbaum$^{1,2}$\\[0.3cm]
 $^1$Institut f\"ur Kernphysik, Forschungszentrum J\"ulich, D-52425 J\"ulich, Germany\\[0.1cm]
 $^2$HISKP (Theorie), Univertit\"at Bonn, Nu\ss{}allee 14-16, 53115 Bonn, 
 Germany
 \end{center}
 
Chiral effective field theory provides a systematic and controlled framework to 
study the dynamics of few-nucleon systems \cite{Weinberg:1991um}. It relies on the low-momentum expansion and 
allows to derive nuclear forces and current operators from the most general 
effective Lagrangian for pions, nucleons and external sources in  
harmony with (approximate) chiral symmetry of QCD. 

The structure of the nuclear force at few lowest orders in the chiral
expansion were discussed and selected applications to few--nucleon systems
were presented, see \cite{Epelbaum:2005pn} for a recent review article.  
The most advanced studies in the two--nucleon sector at
next--to--next--to--next--to--leading order 
(N$^3$LO) in the chiral expansion demonstrate the ability of 
the chiral forces to provide an accurate description of the data in the
low--energy region \citetwo{Entem:2003ft}{Epelbaum:2004fk}.
For three-- and more nucleon systems calculations have so far been performed
up to next--to--next--to--leading order (NNLO). At this order one has to
take into account for the first time the chiral three--nucleon force (3NF) 
\citetwo{vanKolck:1994yi}{Epelbaum:2002vt}. Most of the calculated elastic and breakup 
nucleon--deuteron (Nd) scattering observables are in a
reasonable agreement with the data up to $E_{\rm lab} \sim 65$ MeV. 
In few cases such as the vector analyzing power in elastic Nd scattering and 
the Nd breakup cross section in certain configurations \cite{Ley:2006hu} large
discrepancies with the data are observed. 

It is important to extend the calculations for three and more nucleons 
to N$^3$LO in order to test the convergence of the chiral expansion 
and to be able to increase the energy
range. This will require the inclusion of the leading corrections to the 3NF 
which have yet to be worked out. Work along these lines is underway. For
four-- and more--nucleon systems, the leading four--nucleon force (4NF) will also
need to be taken into account. The parameter--free expressions
for the 4NF at N$^3$LO have recently been derived  \cite{Epelbaum:2005zc}.

\newabstract %4 Rusetsky
\label{abs:Rusetsky}

\begin{center}
{\large\bf The theory of pionic deuterium: status and perspectives} \\[0.5cm]
Ulf-G. Mei\ss ner$^{1,2}$, Udit Raha$^1$ and {\bf Akaki Rusetsky}$^{1,3}$\\[0.3cm]
$^1$HISKP (Theorie), Univertit\"at Bonn, Nu\ss{}allee 14-16, 53115 Bonn,
Germany\\[0.1cm]
$^2$Forschungszentrum J\"ulich, Institut f\"ur Kernphysik (Theorie), D-52425 J\"ulich, Germany\\[0.1cm]
$^3$On leave of absence from: High Energy Physics Institute, Tbilisi State University, 0186 Tbilisi, Georgia
\end{center}

In this talk I give a comprehensive survey of theoretical calculations
of the pion-deuteron scattering length in chiral effective field theories
-- both in the isospin-conserving and in the isospin-breaking sectors.
Namely, it is demonstrated that the estimated systematic uncertainties 
in the isospin-conserving part of the pion-deuteron scattering length can not
be responsible for the huge discrepancy, which emerges in the
recent analysis of the 
pionic hydrogen and pionic deuterium data from Pionic Hydrogen
collaboration at PSI. If has been further demonstrated
that isospin-breaking corrections to the pion-deuteron
scattering length can be very large, because of the vanishing of the
isospin-symmetric contribution to this scattering length at leading order
in chiral perturbation theory. What is most interesting, these corrections
can explain the bulk of the above-mentioned discrepancy.
We also give the first estimate
of the size of the electromagnetic low-energy constant~$f_1$.

Further, we propose to include the correlations, which are due to the 
presence of the same low-energy constants in different bound-state
observables, in the simultaneous analysis of
the pionic hydrogen and pionic deuterium data. In this manner, one may
substantially reduce the systematic error, arising from these low-energy 
constants and determine the $S$-wave $\pi N$ 
scattering lengths at a much better accuracy. To this end, however,
one needs to evaluate the isospin-breaking corrections at least at
order $p^3$ both in the pionic hydrogen and pionic deuterium.

The main results of the talk are contained in 
Refs.~\citetwo{Meissner:2005ne}{Meissner:2005bz}.

\newabstract %5 Baru
\label{abs:Baru}
 \begin{center}
 {\large\bf ChPT for $NN\to NN\pi $ and absorption correction to
$a_{\pi d}$ 
}\\[0.5cm]
 {\bf V. Baru}$^1$, C. Hanhart$^2$, J. Haidenbauer$^2$,
 A. Kudryavtsev$^1$, \\V. Lensky$^{1,2}$ and U.-G. Mei\ss ner$^{2,3}$\\[0.3cm]
{\small $^1$ Institute of Theoretical and Experimental Physics, %\\
117259, \\ B. Cheremushkinskaya 25, Moscow, Russia} \\
{\small $^2$ Institut f\"{u}r Kernphysik, Forschungszentrum J\"{u}lich GmbH,}%\\ [0.3cm]
{ D--52425 J\"{u}lich, Germany} \\
{\small $^3$ Helmholtz-Institut f\"{u}r Strahlen- und Kernphysik (Theorie), 
} \\ 
{\small Universit\"at Bonn, Nu{\ss}allee 14-16, D--53115 Bonn, Germany 
}
% $^1$Laboratori Nazionali di Frascati dell' INFN, C.P. 13-I-00044, Frascati, 
% Italy\\[0.1cm]
% $^2$HISKP (Theorie), Univertit\"at Bonn, Nu\ss{}allee 14-16, 53115 Bonn, 
% Germany\\[0.1cm]
% $^3$Stefan Meier Institut f\"ur subatomare Physik, \"Osterreichische Akademie 
% der Wissenschaften, Boltzmangasse 3, A-1090, Wien, \"Osterreich
 \end{center}

We present the parameter free calculation for the reaction $NN\to
NN\pi $ up to NLO in ChPT \cite{NNpi} utilizing the counting scheme that
acknowledges the large center-of-mass momentum $p\sim
\sqrt{m_{\pi}M_N}$ between initial nucleons (see review \cite{report} and Refs.
therein).  It turns out that in this counting scheme some loops start 
to contribute already at NLO. 
Moreover the  contribution of these loops to the amplitude of the reaction $pp\to
d\pi^+$ at NLO grows linearly with respect to the final NN
relative momentum. This behavior leads to a large sensitivity to the
final NN wave function when the convolution integral with the
transition operator is evaluated. However the central finding of our
recent paper \cite{NNpi} is that there are additional irreducible terms that
contribute at     the same order and cancel exactly the linear growth
to restore the consistency of the formalism.  These terms stem from
the so called box diagrams that allow the two-nucleon intermediate
cuts and therefore are formally reducible  (see Fig. 1a) in Ref
\cite{NNpi}).  However the part of
 the Weinberg-Tomozawa $\pi N\to \pi N$ vertex cancels  one of the nucleon
propagators. This produces additional irreducible terms that bring the
sum of all loops at NLO to vanish.
Moreover the rest of the box diagrams after a proper separation of the
irreducible terms is a purely reducible contribution with one important
modification as compared to the standard treatment, namely that the $\pi N \to
\pi N$  vertex is to be on shell ($2 m_{\pi}$ instead of standard
$3/2 m_{\pi}$). This enhancement by a factor of 4/3 for the 
amplitude is sufficient to bring the cross section for $pp\to d\pi^+$
close to the experiment.  

Once the reaction $pp\to d\pi^+$ is understood within ChPT one can
apply the formalism and the counting rules for the calculation of the
dispersive and absorptive corrections to the $\pi d$ scattering
length. We have recently done this work \cite{pid}. Specifically we
have calculated both the hadronic $\pi d\to NN\to \pi d$ and
the photonic $\pi d\to \gamma NN\to \pi d$ processes . We have shown that
as soon as all diagrams at leading order are included their net effect
on the real part of $a_{\pi d}$ is negligible. 
%This happens due to a
%significant cancellation of the diagrams containing at most one pion
%in the intermediate state (those are also into account in the Faddeev
%calculations) with the so called crossed diagrams. 
Also we get that
both the ratio and the sum of the hadronic and photonic absorptive
corrections to  the imaginary part of  $a_{\pi d}$  are in agreement with the data.

\newabstract %6 Itahashi
\label{abs:Itahashi}
\begin{center}
{\large\bf Future programs for the precision spectroscopy of pionic atoms in the nuclear reactions}\\[0.5cm]
{\bf K. Itahashi}$^1$, R.S. Hayano$^2$, M. Iwasaki$^1$, P. Kienle$^3$, H. Outa$^1$, and K. Suzuki$^3$\\[0.3cm]
{\small 
$^1$ Advanced Meson Science Laboratory, RIKEN, 2-1 Hirosawa, Wako-shi, Saitama 351-0198, Japan\\[0.1cm]
$^2$ Department of Physics, University of Tokyo, 7-3-1 Hongo, Tokyo 113-0033, Japan\\[0.1cm]
$^3$ Physik-Department, Technische Universit\"{a}t M\"{u}nchen, D-85748 Garching, Germany\\[0.1cm]}
\end{center}

A recent technical break-through of pionic atom spectroscopy has been yielding 
precious information on the pion$-$nucleus strong interaction~\cite{PRL72302}, 
and stimulates theoretical investigations on the origin of hadron mass in the
viewpoint of the restoration of the Chiral symmetry in nuclei~\cite{Weise}.

Experimentally, the precision spectroscopy of deeply bound pionic atoms 
is one of the most important achievements, and we have employed 
recoil-free $(d,^3{\rm He})$ nuclear reactions to form the pionic atoms
and performed precision spectroscopy. The precision of the measured
binding energies and widths for the pionic $1s$ tin (Sn 115$\sim$123) isotopes go down 
to $\sim$ 20 keV and 80 keV, respectively.

Presently, the method is well established and 
this field of spectroscopy is approaching the next steps: we have
two directions. One is to perform systematic measurement covering 
wide range of pionic atoms with highest possible resolution.
Another is to explore possibility of making spectroscopy on the
pionic atoms with unstable nuclei, which still requires
basic instrumental studies. 

What we need to focus in the near future is the former, the systematics
study of pionic atoms with wide range of nuclei. This requires
many indispensable properties in the experimental facility.
For instance, the accelerator is expected to provide 250 MeV/nucleon
deuteron beam with higher intensity than $1 \times 10^{12}$/sec.
This is because ... the formation cross section of the pionic atom in 
the nuclear reaction is not larger than several tens of micro barn per
steradian. Since target thickness is one of the largest factor to
determine the experimental resolution, there is no way to choose
thick target...

After examining the possibility of performing the systematic study 
on the accelerator facilities in the world, we came to conclude that
the newly-built facility in RIKEN, RI beam factory (RIBF), is
most suited for this purpose. The facility has an accelerator
complex to provide very high quality deuteron and 
heavy ion beams with sufficiently
high intensity and high duty factor. The projectile fragment separator,
BigRIPS will provide an excellent performance as a spectrometer with
its flexible optical settings. The first beam is scheduled in the
year 2007. We will just start the practical preparation for the experiment now.
Any contributions are welcomed.

\newabstract %7 Borasoy
\label{abs:Borasoy}
\begin{center}
{\large\bf Low-energy \boldmath{$\bar{K} N$} interactions}\\[0.5cm]
B.~Borasoy\\[0.3cm]
Universit\"at Bonn,\\
Nu{\ss}allee 14-16, 53115 Bonn,
Germany\\[0.1cm]
\end{center}

The low-energy $\bar{K}N$ system is of special interest as a testing ground for
chiral SU(3) symmetry in QCD and, in particular, for the role of explicit symmetry
breaking induced by the relatively large mass of the strange quark.
Most significantly, the existence of the $\Lambda(1405)$ resonance just  25 MeV
below the $K^- p$ threshold makes chiral perturbation theory inapplicable in this channel.
Non-perturbative coupled-channel techniques based on driving terms of the chiral SU(3)
effective Lagrangian have proved useful and successful in dealing with this problem,
by generating the $\Lambda(1405)$ dynamically as an I = 0  $\bar{K}N$ quasibound state
and as a resonance in the $\pi\Sigma$ channel. High-precision $K^- p$ threshold data set
important constraints for such theoretical approaches.

In several studies \citefour{BNW1}{BNW2}{BNW3}{BMN} we have investigated the $K^- p$
system within different variants of chiral unitary approaches.
Based on a very large variety of different fits to data we can provide an error range
for the strong $K^- p$ scattering length
which is related to the strong interaction shift and width in kaonic hydrogen.
We obtain an energy shift and width in kaonic hydrogen which is in agreement
with the KEK experiment, but disagrees with DEAR.
Our analyses point on questions of consistency of the
recent DEAR measurement with previous $K^- p$ scattering data.
The conservative
error range for $a_{K^- p}$ derived from chiral unitary approaches is in clear
disagreement with the one deduced from the DEAR experiment.

The upcoming measurement at SIDDHARTA@DA$\Phi$NE aiming for a precision at the level
of a few electron volts in the shift and width of kaonic hydrogen
will further clarify the situation.
Our investigations are also of considerable interest in the discussion
of possible deeply bound $K^-$-nuclear states.

\newabstract %8 Oller
\label{abs:Oller}
\begin{center}
{\large\bf About the Strangeness $-1$ S-wave Meson-Baryon Scattering}\\[0.5cm]
Jos\'e A. Oller\\[0.3cm]
Departamento de F\'{\i}sica, Universidad de Murcia, E-30071 
Murcia, Spain
\end{center}

We consider meson-baryon interactions in S-wave with strangeness $-1$. This is a 
non-perturbative sector
populated by plenty of resonances interacting in several two-body coupled channels.  We
study this sector combining a large set of experimental data.  The recent experiments 
from the Crystal Ball Collaboration are  remarkably accurate demanding a
sound theoretical description to account for all the data. We employ unitary chiral
perturbation theory   to accomplish this aim \citetwo{mop}{o}. The approach 
is employed up to and including ${\cal{O}}(p^2)$ baryon CHPT amplitudes which are then 
used as interaction kernels in a general expression that resums the right hand or unitarity
cut, making use of unitarity plus analiticity.  We find two types of solutions that 
agree well with scattering data. However, while one of these solutions, the so called 
solutions A, agree well with the accurate measurement by the DEAR Collaboration of the
width and shift of the energy of the fundamental state of kaonic hydrogen this is not 
the case with other solutions, the so called solutions B.   
 The spectroscopy of our solutions is studied in detail in ref.\cite{o}, 
 discussing  the rise 
 from the pole content of the two $\Lambda(1405)$ resonances,  $\Lambda(1670)$, 
 $\Lambda(1800)$, $\Sigma(1440)$, $\Sigma(1620)$ and
 $\Sigma(1750)$ for the solutions of type A. Notice that all these resonances are the 
 ones appearing in the PDG from $\pi\Sigma$ threshold up to 1.8 GeV, with the 
quantum numbers we are considering. However, B type solution is only able to generate 
the two $\Lambda(1405)$, $\Lambda(1670)$ and $\Sigma(1440)$ resonances.
 We also argue about the fact
that there are two I=1 poles before and above the $\bar{K}N$ threshold that finally give
rise to bumps in the physical amplitudes with an effective width of around 20-25 MeV,
although the imaginary part of their poles positions are smaller than the half of this 
width. Since they appear consecutively before and above threshold this is why 
the resulting width is 
significantly larger. We also show that the $\Sigma^+\pi^-$ and $\Sigma^-\pi^+$ event 
distributions 
measured in ref.\cite{hemingway} show different shapes, with a $\Lambda(1405)$ peak 
clearly displaced from one another, that can be naturally explained if one allows for 
I=1 resonances around the $\bar{K}N$ threshold. We also consider other parametrizations
with a constant term for the I=1 amplitude for $\pm 100$ MeV around the $\bar{K}N$ 
threshold and the reproduction of these data is much worse. This is considered as an 
evidence for the presence of the I=1 poles at these energies. Since the solutions 
of type A offers a very good agreement with scattering, kaonic hydrogen and 
spectroscopy present data, including DEAR measurement,
 we prefer them over the solutions of type B, which  do not agree 
with DEAR and do not give rise to so many resonances.

\newabstract %9 Nissler
\label{abs:Nissler}

\begin{center}
{\large\bf Chiral unitary approach to \boldmath{$K^- p$} scattering}\\[0.5cm]
R.~Ni{\ss}ler\\[0.3cm]
HISKP (Theorie), Universit\"at Bonn,\\
Nussallee 14-16, 53115 Bonn,
Germany\\[0.1cm]
\end{center}

Chiral unitary approaches combine Chiral Perturbation Theory, the low-energy effective field 
theory of QCD, with non-perturbative methods based on unitarity of the $S$-matrix 
and have been very successful in describing meson-meson and meson-baryon processes at 
low energies---in particular in the vicinity of resonances. 
However, in most such calculations found in the literature an examination of theoretical 
errors has not been undertaken. Based on previous work \citethree{BNW1}{BNW2}{BNW3} we provide
a thorough investigation of theoretical uncertainties within the framework of chiral 
unitary approaches for $K^- p$ scattering \cite{BMN}.

One main source of theoretical errors is related to different choices of input from the 
chiral Lagrangian, e.g., some authors prefer to work with the leading order chiral Lagrangian while 
others also take next-to-leading order terms into account. In order to estimate the pertinent
uncertainty, we compare the results obtained with three different interaction kernels derived
from the leading- and next-to-leading order Lagrangian.

The unknown parameters of the approach are constrained by performing a least-squares fit to 
$\bar{K} N$ scattering data. Making use of Monte-Carlo methods we have performed an extremely 
large number of fits and estimated the statistical errors which result from the fitting procedure.
We can thus provide 1$\sigma$ confidence regions for our theoretical results.

Furthermore, we have critically investigated the pole structure of the fits, in particular
the isospin zero poles in the energy region of the $\Lambda(1405)$ resonance.
We have also illustrated that the general pole structure of a fit
can serve as a criterion to consider the fit as unphysical.
In this respect, the upcoming $\Lambda(1405)$ electroproduction experiment at the ELSA 
accelerator in Bonn may help to further clarify the pole
structure of the $K^- p$ scattering amplitude below threshold.

\newabstract %10 Marton
\label{abs:Marton}
 \begin{center}
 {\large\bf Kaonic Hydrogen Experiments}\\[0.5cm]
 J. Marton for the DEAR/SIDDHARTA Collaborations\\[0.3cm]
 Stefan Meyer Institut f\"ur subatomare Physik, \"Osterreichische Akademie
 der Wissenschaften, Boltzmangasse 3, A-1090, Wien, \"Osterreich
 \end{center}

 Research on kaonic atoms using X-ray spectroscopy is conducted by
 the DEAR/ SIDDHARTA Collaborations at LNF-INFN, Frascati, Italy. The kaonic
 hydrogen atom represents the simplest hadronic atom with
 strangeness. The strong interaction kaon-proton leads to the
 energy shift $\epsilon_{1s}$ and width  $\Gamma_{1s}$ of the 1s state.
 After the KEK experiment \cite{masa1}
 identified the repulsive-type character of the  K$^{-}$p interaction, the DEAR experiment
 verified this finding but found smaller values for $\epsilon_{1s}$ and
 $\Gamma_{1s}$ with higher precision
 \cite{beer05}. These results stimulated new theoretical work
 \cite{theory1}.
 The future of such type of measurements is related to the development of
 new large area X-ray silicon drift detectors (SDDs), conducted by
 the SIDDHARTA Collaboration \footnote{Work supported by EU within
 I3-HadronPhysics and TARI-INFN, Contract
 No. RII3-CT-2004-506078.}. SDDs will provide the application
 of the time correlation between the emitted charged kaon pair in
 DAFNE (where the measurement will be performed) and the kaonic
 hydrogen X-ray detection. In comparison with DEAR, the
 signal-to-background ratio will be improved by 2-3 orders of
 magnitude. A precision
 for kaonic hydrogen $\epsilon_{1s}$ and $\Gamma_{1s}$ at the percent level is
 anticipated (see fig.1). Moreover, for the first
 time the X-ray spectrum of kaonic deuterium will be measured.

\begin{figure}[h]
\centerline{\includegraphics [scale=0.2] {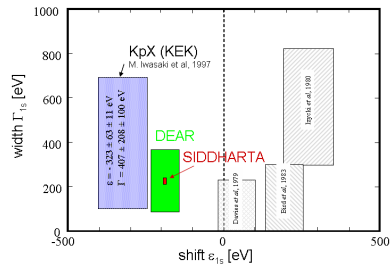}}
  \caption{Experimental values for the kaonic hydrogen strong interaction shift and width.
  The anticipated precision using SDDs is seen as small rectangular area.  }
 \label{fig:2}
\end{figure}

\newabstract %11 Yazkov
\label{abs:Yazkov}
\begin{center}

{\large\bf Measurement of the $\pi^+\pi^-$ atom lifetime at DIRAC}\\[0.5cm]
V. Yazkov$^1$ on behalf of DIRAC Collaboration\\[0.3cm]
$^1$Skobeltsyn Institute of Nuclear Physics of Moscow State University,\\ 
1 - 2, Leninskie Gory, GSP-2, Moscow, 119992, Russia \\[0.1cm]
\end{center}

Pionium or $A_{2\pi}$ is a hydrogen-like atom consisting of $\pi^+$ and 
$\pi^-$ mesons. The goal of the DIRAC experiment at CERN (PS212) is 
to measure the pionium lifetime with $10\%$ precision. Such a measurement
would yield a precision of $5\%$ on the value of the $S$-wave
$\pi\pi$ scattering lengths combination $\left|a_0-a_2\right|$ which is
predicted to be $a_0-a_2=0.265\pm0.004$ \cite{Colangelo:2001}. 
Corresponded $A_{2\pi}$ lifetime is 
$\tau = (2.9 \pm 0.1) \cdot 10^{-15}$~s~\cite{Gasser:2001}.

The $A_{2\pi}$ are produced by Coulomb interaction in the final state 
of $\pi^+\pi^-$ pairs generated in proton--target interactions 
\cite{Nemenov:1985}. Some of them are broken up due to their interaction 
with matter of a target producing 
``atomic pairs'', characterized by small pair c.m. relative momenta 
$Q < 3$~MeV/$c$. Other atoms annihilate into $\pi^0\pi^0$. The amount of 
broken up atoms $n_A$ depends on the lifetime which defines the decay rate. 

Also $\pi^+\pi^-$ pairs are generated in free state. Essential fraction of 
such pairs (``Coulomb pairs'') are affected by Coulomb interaction. 
Number of generated atoms ($N_A$) is proportional to a number of 
 ``Coulomb pairs'' ($N_A=K \cdot N_C$). The coefficient $K$ is precisely 
calculable. The dependence of breakup probability  
$P_{\rm br}(\tau)=n_A/N_A=n_A/(K\cdot N_C)$ on the lifetime $\tau$ is 
determined by the solution of differential transport equations 
\cite{Afanasyev:1996}.

The DIRAC experiment uses a magnetic double-arm spectrometer at the
CERN 24~GeV/$c$ extracted proton beam T8 \cite{Adeva:2003}. 
The data used for this work were taken in 2001 with Ni target 
($\sim 42$~\% of data).
The number of ``atomic pairs'' $n_A=6530 \pm 294$ is obtained as 
an excess of experimental distribution above the approximated 
distribution of ``free pairs'' only in the region $Q<4$~MeV/$c$. 
It provides break up probability 
$
    P_{\rm br}=0.452\pm
    0.023_{stat}~^{+0.009}_{-0.032}\}_{syst}=0.452~^{+0.025}_{-0.039}
$.

Taking into account the dependence of breakup probability on lifetime 
\cite{Afanasyev:1996} a first result on the pionium lifetime is~\cite{Adeva:2005}:

\begin{equation}
    \tau_{1S}=\left[2.91~^{+0.45}_{-0.38}\}_{stat}~^{+0.19}_{-0.49}\}_{syst}\right]\times
    10^{-15} ~\mathrm{s}=\left[2.91~^{+0.49}_{-0.62}\right]\times
    10^{-15} ~\mathrm{s}.
    \label{eq:tauresult}
\end{equation}

This lifetime corresponds to 
$\left|a_0-a_2\right|=0.264~^{+0.033}_{-0.020}~m_{\pi}^{-1}$.

\newabstract %12 Raha
\label{abs:Raha}
\begin{center}
{\large\bf $\mathbf{\bar{K}N}$ Scattering Lengths From Kaonic Deuterium} \\[0.5cm]
Ulf-G. Mei\ss ner$^{1,2}$, {\bf Udit Raha}$^1$ and Akaki Rusetsky$^{1,3}$
\\[0.3cm]
$^1$HISKP (Theorie), Univertit\"at Bonn, Nu\ss{}allee 14-16, 53115 Bonn,
Germany
\\[0.1cm]
$^2$Forschungszentrum J\"ulich, Institut f\"ur Kernphysik (Theorie), 
D-52425 J\"ulich, Germany 
\\[0.1cm]
$^3$On leave of absence from: High Energy Physics Institute, Tbilisi State University, 0186 Tbilisi, Georgia
\end{center}

    My talk deals with the extraction of the $S$-wave $\bar K N$
    scattering lengths from a combined analysis of experimental data
    on kaonic hydrogen and kaonic deuterium which is important in the
    context of the recently proposed SIDDHARTA collaboration expriment
    at Frascati. While still awaiting the first results from the
    SIDDHARTA collaboration, we present a systematic study of the
    near-threshold kaon-deuteron scattering within the framework of
    low-energy effective field theory, using the existing DEAR or KEK
    data for kaonic hydrogen and ``synthetic'' experimental data for
    kaonic deuterium. In our analysis, we consider the partial
    re-summation of the multiple scattering series for the $K^- d$
    scattering in the Fixed Center Approximation Limit. In this limit,
    we show that the isospin breaking effects in the $Kd$ system are
    small, in-spite of the large unitary cusp corrections in the the
    $\bar K N$ system. What is more interesting is that with the
    present DEAR central values of the width and energy shift of the
    kaonic hydrogen ground-state, very stringent constraints have to
    be imposed on the input $Kd$ scattering length in order to ensure
    that the physical solutions for the $\bar K N$ scattering lengths
    $a_0$, $a_1$ exist. In case of the KEK data such constraints turn
    out to be much milder.

The main results of the talk are contained in 
Refs.~\cite{Meissner:2006}.

\newabstract %13 Trautmann
\label{abs:Trautmann}
\begin{center}
{\large\bfseries How accurate are the pionium breakup calculations?}\\[0.5cm]
{\bfseries D. Trautmann}$^1$, T. Heim$^1$, K. Hencken$^1$, and G. Baur$^2$\\[0.3cm]
$^1$Institut f\"ur Physik, Universit\"at Basel, Klingelbergstr. 82, 4056
Basel, Switzerland\\[0.1cm]
$^2$Institut f\"ur Kernphysik, Forschungszentrum J\"ulich, 
52425 J\"ulich, Germany
\end{center}
Our group  provides very precise 
calculations of electromagnetic excitation and ionization cross sections for 
pionium as required for the analysis of
the experiment DIRAC. In order to achieve the required accuracy of 
1\% for the electromagnetic cross sections we have to employ sophisticated 
atomic scattering models taking into account even minor effects of the 
interaction of pionium atoms with the target material. 

Our calculations incorporate a fully quantum mechanical treatment 
of the electromagnetic transitions in pionium  \cite{Halabuka:1999};
target-elastic (coherent) as well as incoherent (target-inelastic) scattering processes
within the framework of Dirac-Hartree-Fock theory 
\cite{Heim:2000};  the explicit determination of magnetic and relativistic terms
in the pionium--atom interaction \cite{Heim:2001}. Our calculated momentum distributions
for the pions from breakup of pionium \cite{Heim:2002} help to reduce the
background in the analysis of the experimental data. Higher order
contributions and multi-photon exchange have been calculated in the Glauber
approximation \cite{Schumann:2002}.
Proceeding from the elementary interaction between pionium and a single target
atom, we have studied the propagation of the pionium atom 
as it moves through the target material with a Monte Carlo approach 
\cite{Santamarina:2003}. The distribution of different states, 
as well as the probability of ionization vs. excitation are of crucial 
importance for the analysis of DIRAC. Since a description in terms of
transition \emph{probabilities} (rather than propagating the \emph{amplitudes})
is not fully justified, we have studied the influence of degenerate states 
within an `optimal mixture' approach \cite{Hencken:2003}. The difference 
within the Monte Carlo simulation was found to be of the order of 0.5\%. 

While our calculations are consistently accurate within our own model to
better than 1\%, we have to allow for the possibility that the target atomic
structure functions (form factors and scattering functions) may present a
more serious limitation of the overall accuracy than previously estimated.
Comparisons of cross section calculations with various atomic form factor
models, including other state-of-the-art calculations 
\citetwo{Hubbell:1975}{Hubbell:1979}
reveal a seemingly systematic shift in the results obtained. A direct 
comparison of these calculations is, however, hampered by several 
circumstances. For one, the tables of factors found in the literature 
have been prepared with the purpose of `conventional' atomic scattering in mind.
Therefore, these tables concentrate on maximum accuracy and reliability in 
a range of momentum transfer that is comparatively small for our setting with
pionium instead of normal atomic scattering. The dominating momentum scale 
relevant for pionium scattering is
typically 136 times larger than that for conventional atomic scattering.
At these high momentum transfers, the existing tables have a less dense grid
of points. Our own calculations of the form factors, on the other hand, put
maximum emphasis on this region of greatest importance for us. 
In addition, we note that complex atomic scattering processes are routinely
calculable only with a limited accuracy, as discussed e.g. in
\cite{Chantler:2000}.

\newabstract %14 Kudryavtsev
\label{abs:Kudryavtsev}
 \begin{center}
 {\large\bf Analytic theory for hadronic atoms beyond the Deser 
approximation}\\[0.5cm]
 A. Kudryavtsev\\[0.3cm]
 ITEP, Bolshaya Cheremushkinskaya 25,~117258, Moscow, Russia \\[0.1cm]
 \end{center}

 The size of lightest hadronic atoms is determined by the Bohr radius 
$a_{B}$. This size is much larger than the range of strong interaction
$r_0$,~$r_0\ll a_{B}$. The spectrum of atom is given by the set 
$\{E_{nl}\}$, where $E_{nl}=E^{Coul}_{nl}+\Delta E^{st}_{nl}$, 
$E^{Coul}_{nl}=-E^C/2n^2$ with $E^C=me^4/\hslash^2$. Here $\Delta 
E^{st}_{nl}$ 
is the
hadronic shift. Usually $\Delta E_{nl}^{st}$
is very small and the Deser approximate equation is valid: 
\begin{equation}
\label{deser}
\frac{\Delta E^{st}_{nl}}{E^C}=\frac{2a^s}{a_Bn^3}~~(l=0)
\end{equation}

This equation works well in the case of the small values of the scattering 
length
$a^s\ll a_B$. However in the case of the strong  attractive  potential 
the 
scattering length $a^s$ may 
become large. In this case the phenomenon of rearrangement of atomic 
spectrum takes place \cite{kumasha}. The equation which describes the 
spectrum for arbitrary values of scattering length was obtained in ref.
\cite{popov}. In the units $\hslash=e=m=1$ it could be written as
\begin{equation}
\label{basic}
2[\psi(1-1/\lambda)+\lambda/2+\ln\lambda]=\frac{1}{a^{cs}}+\frac{1}{2}r^{cs}\lambda^2
\end{equation}
Here $\psi(z)=\Gamma^{\prime}(z)/\Gamma(z)$ and $E=-\frac{\lambda^2}{2}$.

Solving this algebraic equation one may express the energies of 
atomic levels 
and
the energy of a loosely bound nuclear level in terms of the 
Coulomb-modified scattering length $a^{cs}$ and the effective range 
$r^{cs}$.
In the limit of small values for $a^{cs}$ the
equation ~(\ref{basic}) reproduces the Deser equation~~(\ref{deser})
plus small corrections. The leading correction to the Deser's 
formula corresponds to the 
substitution $a^s\Rightarrow a^{cs}$ in the equation (1). The relation 
between
$a^{cs}$ and $a^{s}$ contains large logarithm ($\sim \ln(a_B/r_0)$). It  
is 
discussed in the 
Handbook "Collision Theory" by M.Goldberger and K.Watson, see also  
refs. (\citetwo{popov}{mur}).

\newabstract %15 Ericson
\label{abs:Ericson}
\begin{center}
{\large\bf Corrections to scattering lengths from hadronic atoms}
\\[0.5cm]
{\bf T. E. O. Ericson}$^1$, A. N.
  Ivanov${^2}$,  B. Loiseau$^{3}$ and S. Wycech$^{4}$\\[0.3cm]
$^1$ Theory Division,  CERN, CH-1211 Geneva 23, Switzerland 
\\[0.1cm] 
 $^2$ Atomic Institute of the Austrian Universities,  A-1040
Wien, Austria 
\\[0.1cm]
$^3$ LPNHE, Univ. P. \& M. Curie, 4 Pl. Jussieu, F-75252 Paris, France 
\\[0.1cm]
$^4$ Soltan Institute for Nuclear Studies,
 PL-00681 Warszawa,
 Poland
\end{center}

The high precision reached in the determination of the strong interaction
energy shift and width in the $\pi ^-p$ atom (0.2\%) raises the question to
which extent the Deser-Trueman relation $\epsilon_{1s} \propto a _{l=0}$ gives
an accurate determination of the hadronic scattering length as well as
question of the nature of the corrections \citetwo{ERI04}{ERI06}.  This
relation is exact to order $\alpha ^2$ in terms of the "Coulomb scattering
length".  Our previous results are now generalized and extended. We show that
on general grounds for {\it any} hadronic atom there are 2 classes of
corrections to order $\alpha ^2$ in the scattering length: the coherent ones
with the system remaining in its ground state and intrinsic corrections
associated with the virtual excitation of the system.  The coherent
contributions describe the effect of the external Coulomb field ($"Z\alpha "$)
with extended charges, i.~e., the kinematical effect of the depth of the
Coulomb potential near the origin as well as the correct initial and final
wave function near the origin. The virtual excitations, on the other hand,
gives rise to  genuine isospin breaking  and  they contribute also in the
absence of Coulomb  scattering.  In the single hadronic channel channel case
the coherent corrections are in the limit of zero-range interactions: 

1. The wave function change at the origin for an extended charge 
$-m\alpha   \langle r\rangle _{em}\, a_{0}$.  

2. The change of scattering energy from the threshold value to kinetic energy
   corresponding to the Coulomb potential at the origin for the extended
   charges such that the threshold expansion gives non-relativistically 

3. A  cusp correction which is very insensitive to the detailed charge
   distribution; it corresponds to a  final state wave function consistent
   with the physical scattering length. 

We show that the zero-range approximation is accurate for s-waves with small
additional corrections for the finite interaction range, while the
corresponding result for higher waves depends on the detailed interplay  of
the interaction and charge range. 

  For the $\pi ^-p$ atom, however,   the largest e.\, m. correction is  due to
  the dispersive effect of  $\pi ^-p\, \to \gamma X$ processes, dominated by
  $X=N,\Delta $ with an important contribution from the latter. 
A comparison  to the results of effective ChPT \cite{GAS02} gives both  an
interpretation of the physics of the chiral constants  and their approximate
value. 
The results have obvious implications for isospin violation in $\pi N$ scattering at threshold.

\newabstract %16 Friedman
\label{abs:Friedman}
\begin{center} 
{\large\bf Kaonic atoms as a starting point for antikaon 
nuclear physics}\\[0.5cm] 
E. Friedman \\[0.3cm] 
The Racah Institute of physics, the Hebrew University, Jerusalem, Israel  
\\[0.1cm] 
\end{center} 
 
Recent experimental evidence on candidates for $\bar K$-nuclear deeply 
bound states in the range of binding energy 
$B_{\bar K} \sim 100 - 200$~MeV 
again highlighted the open question of how attractive the $\bar K$-nucleus 
interaction is below the $\bar K N$ threshold, 
a topic which was discussed over a decade ago \cite{BFG97}. The best 
source of information on the antikaon-nucleus potential at threshold 
are strong interaction effects in kaonic atoms. These, however, are 
sensitive to the surface region of the nucleus and the continuation of  
the potential into the nuclear interior is an open question: 
whereas chirally-motivated potentials \cite{CFG01} lead to `shallow' 
real potentials of about 55 MeV deep, various phenomenological approaches 
lead to 180 MeV deep potentials and to {\it significantly better} fits to 
 data. 
The systematics of the kaon-nucleus potential  
 is discussed, particularly in connection with 
the very recent calculations of $\bar K$ nuclear bound states within 
a dynamical model \cite{MFG06}. Figure 1 shows the phenomenological 
antikaon-nucleon interaction density-dependence  
when moving from the interior  
towards  50\% 
and 10\% of the central density (marked with vertical dotted lines). 
 
 \begin{figure}[h] 
 \begin{center} 
 \includegraphics[width=8.5cm]{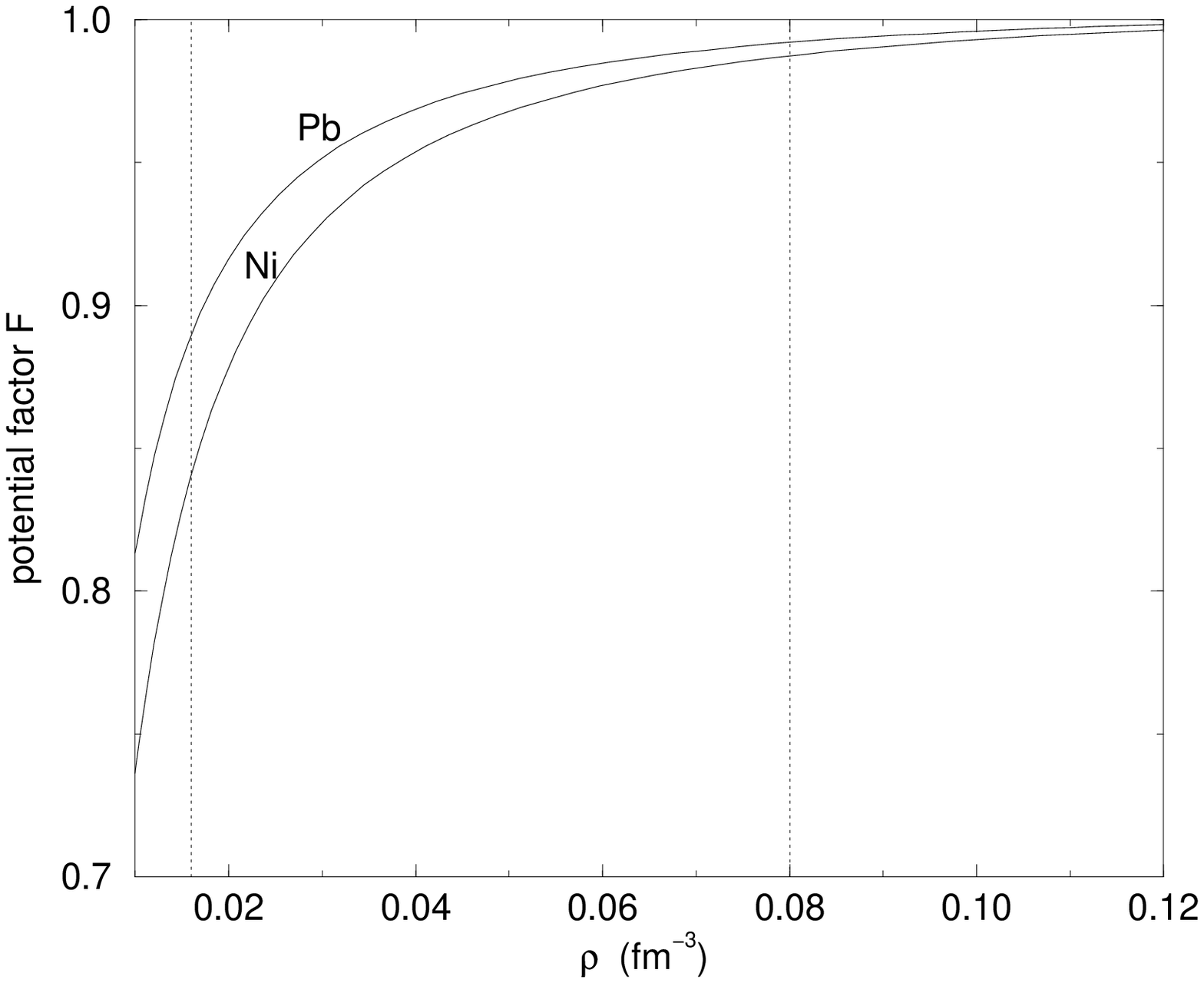} 
 \end{center} 
 \caption{Density-dependence of the antikaon-nucleon interaction.} 
 \label{fig1} 
 \end{figure} 
 
\noindent 
This work was supported in part by the Israel Science Foundation grant 757/05.

\newabstract %17 Hayano_helium
\label{abs:Hayano_helium}
\begin{center}
{\large\bf Determination of the Antiproton-to-Electron Mass Ratio 
by Precision Laser Spectroscopy of $\bar{p}$He$^+$}\\[0.5cm]
R.S.~Hayano (CERN ASACUSA collaboration)\\[0.3cm]
{\em Department of Physics, University of Tokyo, Bunkyo-ku, Tokyo 113-0033, Japan }\\
\end{center}

A femtosecond optical frequency comb and continuous-wave pulse-amplified laser were used to 
measure twelve transition frequencies of antiprotonic helium (metastable
three-body system consisting of an antiproton, an electron and a helium
nucleus)\cite{pbhe} to fractional precisions of  
$(9 - 16)\times 10^{-9}$\cite{hori06}. 
One of these is between two states having microsecond-scale lifetimes hitherto unaccessible to our 
precision laser spectroscopy method. Comparisons with three-body QED calculations yielded an 
antiproton-to-electron mass ratio of $M_{\bar p} /m_e$ =1836.152674(5).  This
also corresponds to a new limit of 2 parts per billion on any possible
difference between the antiproton and proton masses and charges.  

%----- FIGURE  -------------------------------------------------------
\begin{figure}[htb!]
\centering
\includegraphics[height=4cm]{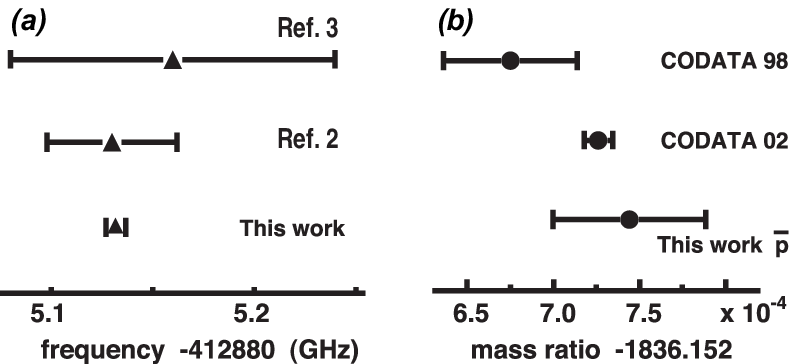}
\caption{\small (a)Frequency of $\bar{p} ^4$He$^+$ transition $(37,35) \rightarrow (38,34)$
measured in this and previous\citetwo{hori03}{hori01} experiments, (b)
proton-to-electron\citetwo{codata98}{codata02} and 
antiproton-to-electron mass ratios.}
\end{figure}

\newabstract %18 Gotta_antiproton
\label{abs:Gotta_antiproton}
 \begin{center}
 {\large\bf Light Antiprotonic Atoms}\\[0.5cm]
D. Gotta\\[0.3cm]
 Institut f\"ur Kernphysik, Forschungszentrum J\"ulich, D-52425 J\"ulich, Germany\\[0.3cm]
 \end{center}
 
 The measurement of the characteristic X--radiation emitted from antiprotonic atoms  
constitutes an antinucleon--nucleus scattering experiment at relative energy zero.  
The strong interaction manifests in an energy shift and broadening of the low--lying  
atomic states, which are directly related to the complex antiproton--nucleus  
scattering length being sensitive to the medium- and long range part of the  
antinucleon--nucleus interaction. The hydrogen isotopes allow access to the elementary  
systems $\bar{p}p$ and $\bar{p}n$. Light nuclei serve as a testing ground to build  
up a consistent picture of the interaction with nuclei\,\cite{Got04}.  
 
In $\bar{p}H$ the resolution of hyperfine states, which is equivalent to a double polarisation  
experiment at threshold, became already possible during the LEAR era\,\citetwo{pbarHK}{pbarHL}.  
However, the low precision -- mainly due to statistics -- hinders a sensitive test of the  
various theoretical approaches. The experimental information on the antiproton-deuteron  
s--wave interaction urgently needs confirmation from a new measurement\,\cite{pbarD} and  
the accuracy of the measurements of the helium isotopes is modest\,\cite{pbarHe}.  
 
For precision studies high statistics is essential. In order to achieve sufficiently large   
X--ray yields antiprotonic hydrogen and helium must be formed in dilute gases to reduce the  
influence of non-radiative de--excitation processes owing to collisions. Therefore, gas  
targets with thin entrance and exit windows must be used. Antiproton beams of 100--300\,keV  
are well suited as planned for the low--energy antiproton facility FLAIR at GSI\,\cite{FLAIR}.  
Combining an antiproton plasma inside a trap with a gas jet might be considered in context  
with the improving performance of such devices. 
 
The low--lying X-ray transitions of hydrogen and helium isotopes are in the energy  
range 2--15\,keV. For hydrogen, the hadronic effects are of the order of 1 keV and 10--500\,meV  
for the s--wave and p--wave interaction, respectively. Consequently, the measurement requires  
two different approaches: a direct measurement with semiconductor detectors, e.\,g., fast CCDs  
and ultimate resolution by using a Bragg crystal spectrometer. Whereas CCDs allow an efficient  
reduction of the annihilation induced background by the analysis of the hit pattern, a Bragg  
spectrometer is self collimating due to the small angular acceptance.

\newabstract %19 Mavromatos
\label{abs:Mavromatos}
\begin{center} 
{\large\bf Looking for ``smoking gun'' signatures of CPT Violation } \\[0.5cm] 
Nick E. Mavromatos,  
\\[0.3cm] 
King's College London, Department of Physics, Strand, London WC2R 2LS,  
U.K. 
\\[0.1cm] 
\end{center} 
 
Due to the weakness 
of the gravitational interaction, one may think that  
prospects of testing quantum aspects of gravity in the  
foreseeable future are futile.  
In this talk I argue that this may not be the case.  
First, I  will present arguments, coming from 
several theoretical approaches to quantum gravity, 
according to which experimental falsification of models within the current or  
immediate future facilitites may have realistic prospects of success.  
One of the most profound aspects of, at least some, models of  
quantum gravity is CPT violation (CPTV)  
``in vacuo'': the latter could either occur  
through spontaneous  
Lorentz violation 
or through quantum decoherence of matter  
propagating in a ``foamy space-time''  
vacuum of quantum gravity~\cite{mavdecoh}. 
In the latter case CPT  is intrinsically violated due to the 
ill-defined  
nature of the CPT operator.  
In the talk I review first some tests associated with  
quantum fluctuations of the space-time metric  
in some models of quantum gravity, 
which lead to the so-called light cone fluctuations~\cite{ford},  
or, equivalently, 
to violations of Lorentz symmetry in individual measurements but possibly 
not on average.  
Then I proceed to discuss some, arguably unique,   
possible signatures of decoherence-induced CPTV 
in entangled particle states~\cite{bmp}, such as neutral mesons in 
meson factories. 
The effects of quantum gravity CPTV in that case (termed $\omega$-effect) 
amount to a  
modification of the pertinent Einstein-Podolsky-Rosen (EPR)  
correlations among  
the neutral mesons produced on each side of the detector.  
They are associated with (direction dependent) interaction  
terms in the part of the meson Hamiltonian  
expressing entanglement  
with the foam. These result  
in modifications of the mass  
eigenstates by the medium of quantum gravity. 
The effects can be (in principle) disentangled experimentally  
from other possible non-unitary effects  
of the foam generated through the evolution of the system in the  
quantum-gravity medium. 
I also present theoretical models of the generation of  
the $\omega$-effect, 
which indicate an order of magnitude  
not far from the experimental sensitivity   
expected to be attained 
at future upgrades of $\phi$-factories such as DA$\Phi$NE.

\newabstract %20 Lehnert
\label{abs:Lehnert}
 \begin{center}
 {\large\bf Lorentz and CPT tests with antimatter}\\[0.5cm]
Ralf Lehnert\\[0.3cm]
Center for Theoretical Physics\\
Massachusetts Institute of Technology, Cambridge, MA 02139, U.S.A.
 \end{center}

One of the most engaging scientific endeavors is 
the search for physics underlying the Standard Model (SM)
and general relativity (GR). 
Substantial theoretical efforts have been devoted to this undertaking. 
Experimental work, 
on the other hand, 
is extraordinarily challenging in this context 
due to the expected Planck suppression of the associated observational signatures. 
However, 
minute Lorentz/CPT violations 
have recently been identified as promising quantum-gravity signals: 
they are amenable to ultrahigh-precision tests, 
and they are predicted by various candidate underlying theories including 
strings, spacetime foam, loop gravity, non-commutative geometry, varying
couplings, and braneworld scenarios.  

Low-energy signatures of Lorentz/CPT violation are described by 
an effective field theory called the Standard-Model Extension (SME). 
This framework incorporates the entire body of established physics 
(i.e., the SM and GR). 
It also includes all Lorentz-/CPT-violating corrections 
compatible with key principles of physics. 
To date, 
the SME has provided the basis 
for numerous studies of Lorentz/CPT breaking 
involving protons, neutrons, electrons, muons, and photons. 
Discovery potential exists in neutrino physics.
 
A particularly promising class of Planck-scale CPT tests 
are matter--antimatter comparisons. 
Note 
that if conventional unitary quantum mechanics remains valid, 
CPT violation implies Lorentz breakdown. 
The SME exemplifies this rigorous result 
and therefore predicts sidereal variations as a general observable feature of CPT breaking. 
(For a model without conventional unitary quantum mechanics, 
see N.~Mavromatos' talk.) 

At present, 
various experimental efforts (ALPHA, ASACUSA, ATRAP) are underway 
to trap cold antihydrogen; 
the goal is to perform hydrogen--antihydrogen spectroscopy 
(see R.~Hayano's and B.~Juhasz' talks). 
The 2-photon 1S--2S transition has received considerable attention 
because an eventual resolution of one part in $10^{18}$ 
seems feasible. 
Each S state contains four levels; 
prior to excitation, 
the $\vec{B}$ field in the trap 
confines the two 1S low--field seekers $|c\rangle_1$ and $|d\rangle_1$. 
Two 2-photon 1S--2S transitions are allowed: 
one between the pure-spin levels $|d\rangle_{1,2}$ 
and one between the mixed-spin states $|c\rangle_{1,2}$. 
The leading-order SME predictions show 
that Lorentz/CPT violation only perturbs the $c$ transition, 
while the $d$ transition is left unaffected. 
Another SME study suggests 
an alternative H--$\overline{\rm H}$ comparison 
employing the 1S Zeeman transition: 
with a $1\;$mHz resolution, 
the competitive bound $|b_3^p|<10^{-18}\;$eV 
on the Lorentz-/CPT-breaking proton parameter $b_3^p$ 
could be achieved. 

Other matter--antimatter Lorentz/CPT tests involve 
proton--antiproton comparisons with Penning traps. 
The SME predicts 
that the anomaly-frequency shifts
are different 
for protons and antiprotons. 
An instantaneous comparison with a $2\;$Hz resolution 
would yield a $10^{-15}\;$eV sensitivity to $b_3^p$. 
Another popular CPT test employs meson interferometry. 
Such measurements are performed 
at the KTeV, OPAL, FOCUS, DELPHI, BELLE, and BaBar experiments. 
Sensitivities of up to $10^{-12}\;$eV 
to the SME quark parameters $\Delta a^{\mu}$ 
have been achieved. 

\newabstract %21 Hayano_antiproton
\label{abs:Hayano_antiproton}
\begin{center} 
{\large\bf Antihydrogen}\\[0.5cm] 
R.S. Hayano\\[0.3cm] 
Department of Physics, University of Tokyo, Bunkyo-ku, Tokyo 113-0033, Japan\\ 
\end{center} 
 
Since motivations to look for CPT-violating effects have been thoroughly
discussed by the previous two speakers\citetwo{Mar}{Leh}, I will first discuss
why we are interested in antihydrogen. High-precision spectroscopy of
antihydrogen is one of the promising ways to test the CPT symmetry, since
methods now exist to measure the $1s-2s$ splitting\cite{Fischer} as well as
the ground-state hyperfine splitting (GSHFS) of ordinary hydrogen to very high
precision (to be discussed by the next speaker\cite{Juh}).  
 
``Cold'' antihydrogen atoms were successfully produced at CERN's antiproton
decelerator (AD) by ATHENA \cite{ATHENA} and ATRAP\cite{ATRAP} collaborations;
both groups used the ``nested Penning trap'' method, in which
synchrotron-radiation-cooled positrons were mixed with electron-cooled
antiprotons. ATHENA's reported {\= H} production rate was about 300 Hz, when
$10^4$ antiprotons were mixed with a positron plasma whose density was $\sim
10^8 {\rm cm^{-3}}$\cite{highrate}.  
 
However, this does not mean that antihydrogen atoms can be readily studied by
high-precision laser spectroscopic methods. The prerequisites are i) atoms are
in the ground state, and ii) atoms are cold (slow) enough so that they stay
long enough in the laser beam.  Although ordinary hydrogen atoms can be cooled
with a cold finger\cite{Fischer}, the method is clearly not applicable to
antihydrogen atoms.  Both ATRAP and ALPHA (successor to ATHENA) groups
therefore plan to trap {\= H} atoms in a minimum-B magnetic trap, whose
typical depth is about 1~K.   
 
At present, there is no positive identification of {\= H} atoms in the $1s$
state; the fact that ATRAP used a field-ionization method to detect {\= H}
atoms indicates that their detected {\=H}s are highly
excited\cite{ATRAP}. Moreover the atoms seem to have much higher energies than
the ambient trap temperature\cite{speed}.   Thus, although antihydrogen-atom
spectroscopy is a promising tool to probe `Planck-scale'
physics\citetwo{Mar}{Leh}, there are still many hurdles to be cleared.

\newabstract %22 Juhasz
\label{abs:Juhasz}

 \begin{center}
 {\large\bf Measurement of the ground-state hyperfine structure of antihydrogen}\\[0.5cm]
 {\bf B.~Juh\'asz}$^1$, and E. Widmann$^1$\\[0.3cm]
 $^1$Stefan Meyer Institut f\"ur subatomare Physik,
Boltzmanngasse 3, A-1090 Vienna, Austria
 \end{center}

The hydrogen atom is one of the most extensively studied atomic systems,
and its ground state hyperfine splitting (GS-HFS) of $\nu_{\mathrm {HFS}}
=1.42$~GHz has been measured with an extremely high precision of
$\delta \nu_{\mathrm{HFS}}/\nu_{\mathrm{HFS}} \sim 10^{-12}$.
Therefore the antimatter counterpart of hydrogen, the antihydrogen
atom, consisting of an antiproton and a positron,
is an ideal laboratory for studying the CPT symmetry.

As a test of the CPT invariance, measuring $\nu_{\mathrm {HFS}}$ of
antihydrogen can surpass in accuracy a measurement of the 1S--2S
transition frequency proposed by other groups. In fact, it has several
advantages over a 1S--2S measurement. Firstly, it does not require the
(neutral) antihydrogen atoms to be trapped. Secondly, the only existing
consistent extension of the standard model, which is based on a microscopic
theory of CPT and Lorentz violation \cite{Blu99}, predicts that
$\nu_{\mathrm {HFS}}$ should be more sensitive to CPT violations.
In addition, the parameters introduced by Kostelecky et al. have
the dimension of energy (or frequency). Therefore, by measuring a
relatively small quantity on an energy scale (like the 1.42~GHz
GS-HFS splitting), a smaller relative accuracy is needed to reach
the same absolute precision for a CPT test. This makes a
determination of $\nu_{\mathrm {HFS}}$ with a relative accuracy of
$10^{-4}$ competitive to the measured relative mass difference of
$K^0$ and $\overline{K^0}$ of $10^{-18}$, which is often quoted
as the most precise CPT test so far.

The ASACUSA collaboration at CERN's Antiproton Decelerator (AD)
has recently submitted a
proposal \cite{ASA05} to measure $ \nu_{\mathrm {HFS}}$ of antihydrogen in
an atomic beam apparatus similar to the ones which were used
in the early days of hydrogen HFS spectroscopy. The apparatus consists
of two sextupole magnets for the selection and analysis of the spin of
the antihydrogen atoms, and a microwave cavity to flip the spin.
This method has the advantage that antihydrogen
atoms of temperatures up to 150 K, ``evaporating'' from a formation
region, can be used. Numerical simulations show that such an
experiment is feasible if $\sim$100 antihydrogen atoms per second
can be produced in the ground state, and that an accuracy of
better than $10^{-6}$ can be reached within reasonable measuring
times.

\newabstract %23 Kuroda
\label{abs:Kuroda}
\begin{center}
{\large\bf MUSASHI -- An ultra-slow antiproton beam source -- \\
Ultra-slow antiproton beam source and antiprotonic atom formation}\\[0.5cm]
{\bf N. Kuroda}$^1$, H.A. Torii$^2$, M. Shibata$^1$, H. Imao$^1$,
 Y. Nagata$^2$, Y. Enomoto$^2$, Y. Kanai$^1$, A. Mohri$^1$,
 K. Komaki$^2$, and Y. Yamazaki$^{1,2}$\\[0.3cm]
$^1$Atomic Physics Laboratory, RIKEN, 2-1 Hirosawa, Wako-shi, Saitama 351-0198, Japan\\[0.1cm]
$^2$Institute of Physics, University of Tokyo, 3-8-1 Komaba, Meguro-ku, \\
 Tokyo 153-8902, Japan\\[0.1cm]
\end{center}

The preparation of large numbers of antiprotons at low energy plays an
important role to synthesize antihydrogen (${\rm \overline{p}e^+}$) and 
antiprotonic atoms (${\rm \overline{p} A^+}$).
Such exotic atoms can only be efficiently synthesized at a few tens of
eV or less.
We, MUSASHI sub-group of ASACUSA collaboration, developed an ultra-slow
antiproton beam source, MUSASHI, Monoenergetic Ultra-Slow Antiproton
beam Source for High-Precision Investigation, with a sequential
combination of the CERN Antiproton Decelerator (AD) and the
radio-frequency quadrupole decelerator (RFQD).
With 10--1000~eV energy beam from MUSASHI, we will study initial process
of antiprotonic atom formation~\cite{Yamazaki:1999} and will make polarized
antihydrogen atoms with cusp trap~\cite{Mohri:2003} for antihydrogen
hyperfine structure measurement~\cite{Eberhard:2001}.

%Antiprotons having GeV scale energy were decelerated to 110 keV by the
%AD and the RFQD. 
The apparatus, MUSASHI, consists of two parts. One is an antiproton
trapping section, so called multiring electrode trap (MRT) housed
in a 4~K bore tube of a 2.5~T superconducting solenoid in which
antiprotons are captured and cooled to sub-eV energy after collisions
between simultaneously confined electrons. 
We succeeded in confining $1.2 \times 10^6$ antiprotons per AD 
shot~\cite{Kuroda:2005}, with $3 \times 10^8$ electrons.
The other part is an ultra-slow antiproton beam transport line~\cite{Ken:2003}.
This beam line has a differential pumping capability for pressure
difference between $10^{-12}$~Torr in the MRT and $10^{-6}$~Torr in the
target gas chamber for antiprotonic atom formation. 
Extracted $3 \times 10^5$ antiprotons with
250~eV energy as an ultra-slow beam were transported to the beam line end.

In the year 2006, we will start our planned experiment, 
${\rm \overline{p}A^+}$ formation cross section measurement with MUSASHI
and a supersonic gas jet target.
At the same time, we installed a new superconducting solenoid cooled by
three refrigerators free of any liquid helium for stable operation and
increasing trapping and extraction efficiency.

\newabstract %24 Wycech
\label{abs:Wycech}
\begin{center}
 {\large\bf On the structure of KNN, KNNN states}\\[0.5cm]
 S. Wycech\\[0.3cm]
  Soltan Institute for Nuclear Studies , Warsaw,
   Poland
 \end{center}

 A semi-quantitative understanding of the KEK~\citetwo{KEK}{KEK1}
 and FINUDA~\cite{FINUDA} findings is attempted. Leaving aside the interpretation
 of the peaks attributed to  Kpp and KNNN systems, two essential  theoretical
 questions arise:

($\bullet$)  what is the binding mechanism, ( $\bullet$)  can the
widths be narrow.

 The K-N scattering amplitudes relevant to  the bound  K-few-nucleon
 systems involve subthreshold energies
determined by $ E_B$ - the kaon and nucleon bindings and $
E_{recoil}$ - the recoil of the KN pair with respect to  the
residual system
\begin{equation}
 \label{label1}
 f_{KN} = f_{KN}( - E_B - E_{recoil})
  \end{equation}
If the binding is as strong as 100 MeV,  the momenta involved in
the wave functions reach 400 MeV/c and the average  recoil energy
amounts to  $ \approx 100$ MeV. The energies of interest $ - E_B -
E_{recoil}$ are located well below the $\Lambda(1405)$ and $
\Sigma(1385)$ states. The amplitudes there are strongly attractive
and give rise to very  strong binding ~\cite{WG}. The energies are
far away from the physical region tested in the KN scattering and
there arise  uncertainties in the KN scattering amplitudes. For
instance, if $\Lambda(1405)$ is a KN bound state, the amplitude
far below the resonance  is given not only by the KN binding but
largely by the Born term which indicates strong dependence on the
uncertain interaction range. Similar dependence occurs with the
$\Sigma(1385)$ resonance.

The K-few-nucleon binding energies are calculated via the
variational procedure with a meson trial wave function generated
with fixed nucleons. This allows to find a contraction potential
in the NN and NNN systems due to the presence of K meson. It
amounts to $\approx 250$ for pp and $\approx 400$ MeV for NNN at
less than 0.5 $fm$ ranges. The real binding is determined by the
repulsive core in the NN interactions. With the  Argonne NN
potential one obtains Kpp state bound by about 50 MeV and KNNN
states bound by about 150 MeV. In the first system the effect of $
\Sigma(1385)$ is small but in the KNNN states it adds large
contribution  to the binding.

The width of about 60 MeV is obtained in the Kpp case as the
result of mesonic decays only. We argue qualitatively that the
widths  in KNNN systems  may be smaller if these are bound by a
150 MeV or more. That comes  as the result of the recoil energy
taken by the spectators in the dominant  non-mesonic decay mode.

\newabstract %25 Gal
\label{abs:Gal}
 \begin{center} 
 {\large\bf Dynamical calculations of $K^-$ nuclear bound states}\\[0.5cm] 
 Avraham Gal\\[0.3cm] 
 Racah Institute of Physics, The Hebrew University, Jerusalem 91904, 
 Israel\\[0.1cm] 
 \end{center} 
 
\begin{figure}[h] 
  \begin{center} 
  \includegraphics[width=6.5cm]{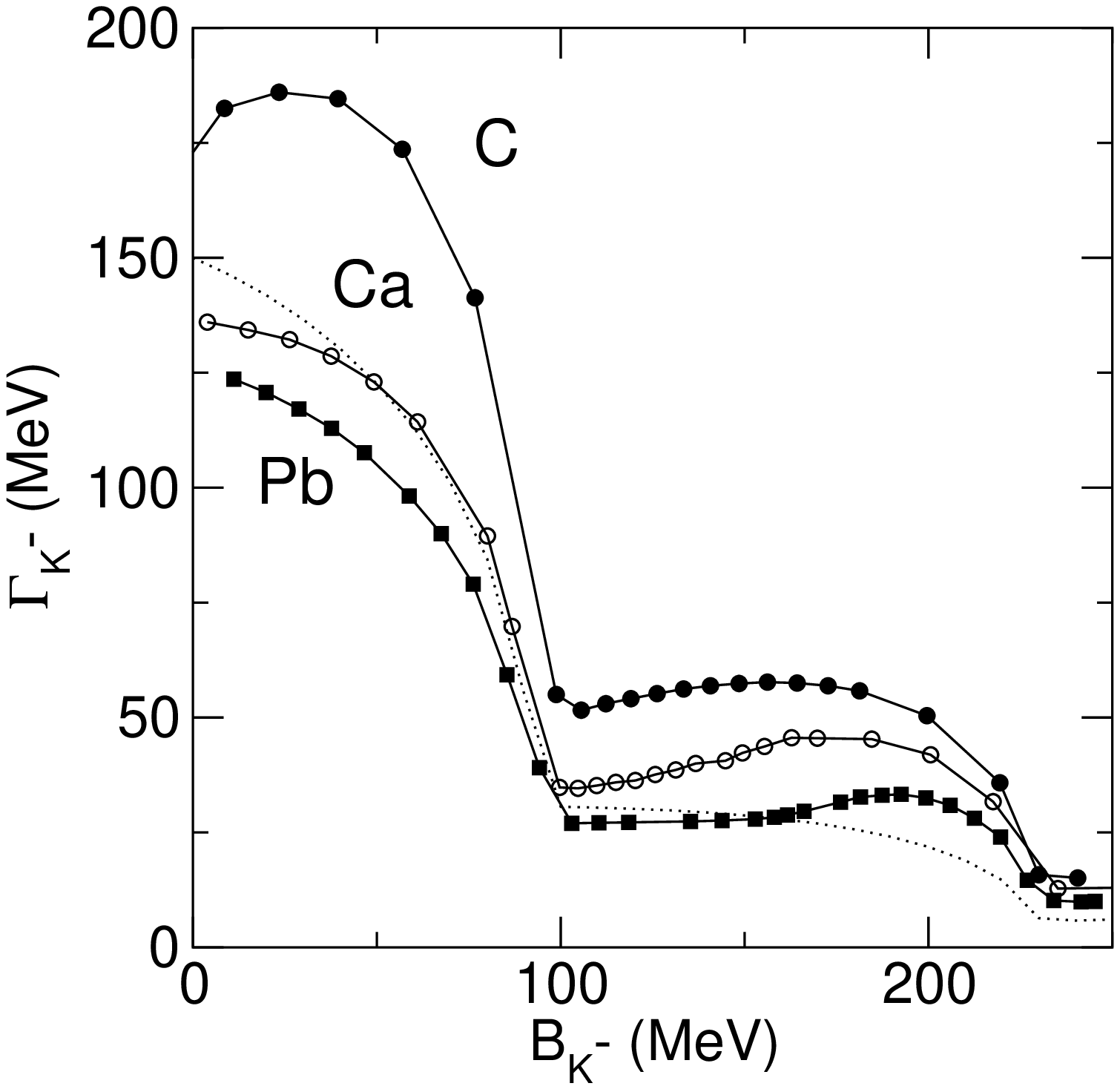} 
  \end{center} 
  \caption{Dynamically calculated widths of $1s$ states 
  as function of the $K^-$ nuclear binding energy for nonlinear RMF models. 
  The dotted line is for a static nuclear-matter calculation with 
  $\rho_0=0.16~{\rm fm}^{-3}$.} 
  \label{fig1gal} 
  \end{figure}

Evidence for and against strongly attractive $\bar K$-nuclear interaction, 
capable of binding $\bar K$s by over 100~MeV where the main decay channel 
$\bar K N \rightarrow \pi \Sigma$ is kinematically closed, is reviewed. 
A key issue is the residual width anticipated from 
$\bar K N N \rightarrow Y N$ absorption modes for $\bar K$ deeply 
bound states. This and other relevant issues in $\bar K$-nuclear dynamics 
are studied, using a relativistic mean-field model Lagrangian 
which couples the $\bar K$ to the scalar and vector meson fields 
mediating the nuclear interactions \cite{MFG05}. The reduced phase space 
available for $\bar K$ absorption from these bound states is taken into 
account by adding a $-{\rm i}s\rho$ imaginary term, with an energy-dependent 
strength $s$ normalized at threshold to fits to the $K^-$ atomic data 
(the subject of $K^-$ atoms is covered by E.~Friedman in these Proceedings). 
$\bar K$-nuclear bound states are generated self consistently over a wide 
range of energies by varying the $\bar K$-meson couplings. Substantial 
polarization of the core nucleus is found in this dynamical model for light 
nuclei, and the binding energies and widths differ appreciably from those 
calculated for a static nucleus. These calculations provide a lower limit of 
$\Gamma_{\bar K} = 50 \pm 10$ MeV on the width of nuclear bound states for 
$\bar K$ binding energy in the range $B_{\bar K} \sim 100 - 200$~MeV, 
as shown in Fig.~\ref{fig1gal}. Comments are made on the experimental signals 
proposed for $\bar K$-nuclear deeply bound candidate states. 
 
\newabstract %26 Oset
\label{abs:Oset}
\begin{center}
 {\large\bf Chiral dynamics of $\bar{K}$-nucleus interaction: critical review
 of deeply bound states}\\[0.5cm]
 {\bf E. Oset$^1$},  V. K. Magas$^2$, A. Ramos$^2$ and H. Toki$^3$\\[0.3cm]
 $^1$Departamento de Fisica Teorica and IFIC, Universidad de Valencia\\[0.1cm]
 $^2$Departament d'Estructura i Constituents de la Materia, Universitat de
 Barcelona\\[0.1cm]
 $^3$Research Center for Nuclear Physics, RCNP, Osaka University
 \end{center}
 
   In this talk I reported on the chiral dynamics of meson baryon interaction
 and present models for the $\bar{K} N$  interaction. Results were reported on
 the selfconsistent evaluation of the $\bar{K}$ selfenergy in a nuclear medium,
 leading to a potential which is in good agreement with $K^- $ atoms.  This
 potential was  shown to be much smaller than the one claimed in the work of 
 Akaishi and Yamazaki (AY), and the differences were traced to several rough
 approximations in the work of AY, like assuming that the $\Lambda(1405)$ is a
 bound state of $\bar{K} N$, when in chiral dynamics is a subtle consequence of
 coupled channel dynamics of  $\bar{K} N$ and $\pi \Sigma$, assuming a Fermi
 sea of infinite nuclear matter for $^3 He$, not including selfconsistency in
 the calculations and compressing the matter to ten times nuclear matter
 density. 
 
    Since the potential obtained by Ramos-Oset (RO), corroborated by other five
independent microscopical studies, leads to deeply bound states with
widths of about 100 MeV, the experimental claims made at KEK and FINUDA of
narrow deeply bound states look incompatible with the chiral predictions. Then
we looked for alternative explanations of the peaks seen at KEK and FINUDA and
we could interpret them in terms of $K^-$ absorption by pairs of nucleons going
to $\Sigma N$ and $\Lambda N$ with no further interaction with the daughter
nucleus in the case of KEK and with final state interaction in the case of
FINUDA. The works are reported in \citetwo{toki}{magas}.  

In the discussion I clarified that the criticism of AY to those works was
unfounded. Akaishi confused the cut off used in chiral theories to regularize
the loop of a meson and a baryon with the range of the potential. These two
concepts have nothing to do with each other. It was also clarified that the
calculation of RO contains the double selfconsistency of using the binding of
the kaons in the intermediate loops and evaluating the selfenergy for kaons with
 the same mass 
that the calculation provides. AY fail to put the selfenergy in the loops which
is fatal in the presence of a resonance, as it is the case here. Yamazaki
claimed that the peak of FINUDA coming from $K^- pp$ absorption did not exist
and I showed that this was not the case, it was obvious in the published paper
and there was a whole paragraph devoted to explain the nature of this peak. I
also clarified that from the  $K^- ~^4 He$ absorption into $\Sigma^- p ~d$, where
the deuteron has about 200 MeV/c from Fermi motion, a peak of strength 1.6
percent measured by Katz (the amount claimed in the KEK experiment) with a width
of about 10  MeV was unavoidable, contrary to what was claimed by Yamazaki.

\newabstract %27 Akaishi
\label{abs:Akaishi}
 \begin{center} 
 {\large\bf Deeply bound kaonic nuclear states in reply to recent criticisms}\\[0.5cm] 
 {\bf Yoshinori Akaishi}$^1$ and Toshimitsu Yamazaki$^2$\\[0.3cm] 
 $^1$College of Science and Technology, Nihon University, Funabashi, Chiba
 274-8501, Japan, and Nishina Center for Accelerator-Based Science, RIKEN,
 Wako, Saitama 351-0198, Japan\\[0.1cm]  
 $^2$Department of Physics, University of Tokyo, 7-3-1 Hongo, Tokyo 113-0033,
 Japan, and Nishina Center for Accelerator-Based Science, RIKEN, Wako, Saitama
 351-0198, Japan\\[0.1cm]  
 \end{center} 
 
The possible existence of few-body kaonic nuclear states was theoretically  
predicted by the present authors \cite{Akaishi02}. Recently, this prediction  
has been critically reviewed by Oset and Toki (OT) \cite{Oset06}.  
Our reply to some of their criticisms is summarized as follows: 
 
1) The phenomenological $\bar K N$ interaction used in our prediction is much  
more attractive than that of Oset-Ramos(OR)'s chiral unitary model  
\cite{Oset98}, nevertheless it is in a reasonable agreement with  
other chiral unitary ones \cite{Kaiser95}  
as judged from forward scattering amplitudes.  
OR's interaction gives unrealistic oscillating behavior  
of the $\Lambda (1405)$ wave function in coordinate space  
due to their sharp-cutoff regularization in momentum space.  
 
2) The double-pole nature of $\Lambda (1405)$ was theoretically discussed  
by Jido {\it et al.}, and its experimental evidence was claimed by  
Magas, Oset and Ramos (MOR) \cite{Magas05}. 
When the residue of a resonance pole is deeply complex,  
the resonance shape and position are largely changed  
by an interference between resonance and background terms.  
MOR's broader peak is explained as a remnant of  
the "narrower resonance pole" due to the interference.  
Thus, MOR's conclusion is not sound.  
The "broader one" of the double poles remains at a relatively narrow width  
of 132 MeV due to unphysical potential barriers coming from  
the sharp-cutoff regularization.  
The "broader resonance pole" effect on $\Lambda (1405)$ is probably  
a kind of artifact. 
 
3) We have formulated a microscopic derivation of the $\bar K$  
optical potential. The optical potential must be distinguished  
between the cases of a $\bar K$ in nuclear bound states and  
of a $\bar K$ in scattering states, which is deep for the former  
and is shallow for the latter.  
The optical potential, as is defined to reproduce not phase shift  
but energy shift for a $\bar K$ in a decaying bound state,  
becomes deep more than 100 MeV attraction.  
OT insists to use a self-consistent "shallow optical potential"  
even for a $\bar K$ in nuclear bound states.  
This opinion, however, has no many-body theoretical foundation.

\newabstract %28 Ivanov
\label{abs:Ivanov}
\begin{center} {\large\bf Phenomenological quantum field theoretic
    model of Kaonic Nuclear Clusters
    $K^-pp$, $K^-pnn$ and so on}\\[0.5cm]
  {\bf A. N. Ivanov}$^{1,2}$, P. Kienle$^1$, J. Marton$^1$, and E. Widmann$^1$\\[0.3cm]
  $^1$Stefan Meyer Institut f\"ur subatomare Physik, \"Osterreichische
  Akademie der Wissenschaften, \"Osterreich\\[0.1cm]
  $^2$Atominstitut der \"Osterreichischen Universitat\"an, Technische
  Universit\"at Wien, \"Osterreich\\[0.1cm]
\end{center}

We propose an oscillator model for the description of the wave
functions of the Kaonic Nuclear Clusters (KNC) $K^-pp$ and $K^-pnn$
observed experimentally ~\cite{FINUDA}. Assuming that all {\it
  stiffnesses} of linear restoring forces are equal we fix the
frequencies of oscillations in terms of the width and binding energy
of the resonance $\Lambda(1405)$ treating it as a bound $K^-p$ state
\cite{Akaishi}.  The binding energy $\epsilon_{K^-X}$ and width
$\Gamma_{K^-X}$ of the KNC $(K^-X)$, where $X = p$, $pp$ and $pnn$,
are defined by \cite{Ivanov}
\begin{equation}\label{label1ivanov}
-\,\epsilon_{K^-X} + i\,\frac{\Gamma_{K^-X}}{2} = 
\int d\tau\,\Phi^*_{K^-X}\,M(K^-X \to K^- X)\,\Phi_{K^-X},
\end{equation}
where $d\tau$ is an element of the phase volume of the system $K^-X$,
$\Phi_{K^-X}$ is the wave function of the KNC $K^-X$ and $M(K^-X \to
K^-X)$ is the amplitude of $K^-X$ scattering. We calculate the
amplitudes of $K^-X$ scattering in the tree--approximation to leading
order in chiral and large $N_C$ expansion within ChPT with non--linear
realisation of chiral $SU(3)\times SU(3)$ symmetry and large $N_C$
expansion. The main contributions to the binding energies come from
the Weinberg--Tomozawa terms. The width of the KNC $(K^-pp)$ is caused
by non--pionic decays only. The main contribution to the width of the
KNC $(K^-pnn)$ comes from the $K^-pnn \to \Sigma^- pn$ decay.

For the binding energies, widths and nuclear matter densities of the
KNC $(K^-pp)$ and $(K^-pnn)$ we obtain the following results
\begin{eqnarray}\label{label2}
\epsilon_{K^-pp}&=& - 118\,{\rm MeV}\;,\; \Gamma_{K^-pp} = 58\,{\rm MeV}\;,\; 
\rho_{K^-pp}(0) = 0.26\,{\rm fm}^{-3},\nonumber\\ 
\epsilon_{K^-pnn}&=& - 197\,{\rm MeV}\;,\; \Gamma_{K^-pnn} = 16\,{\rm MeV}\;,\; 
\rho_{K^-pnn}(0) = 0.53\,{\rm fm}^{-3}.
\end{eqnarray}
The obtained results agree well with the experimental data
~\cite{FINUDA}: $\epsilon_{K^-pp} = -\,115^{+\,6}_{-\,5}\,{\rm MeV}$,
$\Gamma_{K^-pp} = 67^{+\,14}_{-\,11}\,{\rm MeV}$ and
$\epsilon_{K^-pnn} = -\,194.0^{+\,1.5}_{-\,4.4}\,{\rm MeV}$,
$\Gamma_{K^-pnn} < 21\,{\rm MeV}$. They are also in qualitative
agreement with those predicted by Akaishi and Yamazaki
~\citetwo{Akaishi}{Andronic}.

\newabstract %29 Magas
\label{abs:Magas}
\begin{center} 
{\large\bf Simulation of the K- nuclear absorption at FINUDA}\\[0.5cm] 
{\bf V.K. Magas}$^1$, E. Oset$^2$, A. Ramos$^1$, H. Toki$^3$\\[0.3cm] 
$^1$ Departament d'Estructura i Constituents de la Mat\`eria, \\
Universitat de Barcelona,
 Diagonal 647, 08028 Barcelona, Spain\\[0.1cm] 
$^2$ Departamento de F\'{\i}sica Te\'orica and IFIC Centro Mixto\\
Universidad de Valencia-CSIC, Institutos de Investigaci\'on de Paterna \\
Apdo. correos 22085, 46071, Valencia, Spain \\[0.1cm] 
$^3$ Research Center for Nuclear Physics, Osaka University,\\
Ibaraki, Osaka 567-0047, Japan
\end{center} 
 
We performed a theoretical simulation of the $K^-$ absorption process in
different nuclei and show that the peak in the $\Lambda p$ spectrum that was
interpreted  
as a deep $K^- pp$ bound state \cite{finuda} corresponds mostly to the process
$K^- pp \rightarrow \Lambda p$ followed by final state interactions of the
produced  
particles with the daughter nucleus \cite{MORT}. 

To reach the former conclusion, computer simulations are made allowing the
stopped kaons in the nucleus to be absorbed by a pair of nucleons of a local
Fermi sea. The nucleon and the $\Lambda$ emitted in the $K^- pp \rightarrow
\Lambda p$  and $K^- pn \rightarrow \Lambda n$  
absorption processes are allowed to re-scatter with other nucleons in the
nucleus leading to nuclear breakup and producing a $\Lambda p$ invariant mass
spectrum with a distinct peak corresponding to one collision. This peak, which
is analogous to the quasi-elastic peak of any inclusive reaction like
$(e,e')$, $(p,p')$ etc, reproduces the experimental peak \cite{MORT}. Another
peak, broader and at smaller energies coming from baryon secondary collisions,
also appears both in our simulation and in the experimental data \cite{finuda}
at the same place and hence, an explanation for the whole experimental
spectrum is found, which does not require to invoke the creation of the $K^-
pp$ bound state. The agreement between our simulations and the experimental
data is very good, giving 
a $\chi^2$ per data point of 1.25 \cite{MORT}.    

We have presented results for $^6Li$, $^7Li$, $^{12}C$, $^{27}Al$ and $^{51}V$
\cite{MORT}, all of them measured by the FINUDA experiment, although the
spectrum was only shown for the combined data of the three lighter nuclei. The
width of the distribution increases slightly with the nuclear mass while the
peak stays in the same location, in accordance with our interpretation of it
as coming from the quasi-elastic processes. Disentangling the spectrum for
each of the nuclear targets used in the FINUDA experiment would be of
particular relevance because a possible interpretation of the data as evidence
of bound $K^-$ nuclear states would unavoidably produce the peak at a
different energy for each nucleus.

\newabstract %30 Kienle
\label{abs:Kienle}
\begin{center} 
 
{\large\bf Probing the Structure of Nuclei Bound by Antikaons}\\[0.5cm] 
 
Paul Kienle$^{1,2}$ 
 
$^1$Stefan Meyer Institute, \"OAW, Wien\\[0.1cm] 
$^2$Technische Universit\"at München 
 
\end{center}

Following a short review of the the present status of the search for deeply  
bound kaonic states in light nuclei, an outlook is given what to do in the  
future to establish this new field of strangeness dynamics in nuclear clusters.   
It is pointed out that for this purpose detectors covering a large solid angle,  
preferably $4\pi$, are needed for the simultaneous study of the formation and decay  
of the systems in an exclusive reaction experiment. A first step towards this  
goal is the use of FOPI at the GSI Darmstadt for the study of strangeness  
containing nuclear states and hyperon resonances in proton and heavy ion  
induced reactions. 
 
Recently we proposed for exclusive reaction studies in the frame work of the  
AMADEUS project at Daphne in Frascati a modified KLOE detector to search for  
kaonic nuclear clusters using stopped $K^-$ induced $p$ and $n$ knock out reactions on  
cryogenic gas targets, such as $^3$He and $^4$He to start with. The KLOE detector  
allows identifying all charged particles including light nuclei and measuring  
their moments and energies with the wanted accuracy, and has in addition the  
capability of measuring the energies of neutrons and $\gamma$-rays, This allows  
performing missing mass spectroscopy in the reaction channels as well as  
invariant measurement of the decays. 
 
With exclusive experiments it will be possible for the first time to measure  
the mass, the total width and the partial widths of all decay channels.  
In three body decays one can measure Dalitz plots and determine from the  
phase space occupation the size, density distribution and the angular momentum  
of the decaying cluster. 
 
Finally a new method is proposed for implanting 2~$K^-$ in a nucleus using  
antiproton annihilation at rest or in flight to search for systems containing  
double strangeness, which are expected to show very high binding energies and  
large densities for which various phase transitions are predicted.

\newabstract %31 Weise
\label{abs:Weise}
 \begin{center}
 {\large\bf Conditions for antikaon-nuclear bound states}\\[0.2cm]
 W. Weise\\[0.1cm]
 Physik-Department, Technische Universit\"at M\"unchen, D-85747 Garching, Germany\\
 \end{center}
 
This presentation summarizes ongoing work \cite{1} examining the conditions
under which $\bar{K}$- nuclear clusters or bound states might exist.  
Our starting point is chiral SU(3) dynamics based on the three-flavour
meson-baryon effective chiral Lagrangian. When combined with coupled-channel
methods, this approach \cite{2} successfully describes s-wave $\bar{K}N$
scattering, the formation of the $\Lambda(1405)$ resonance and the couplings
to $\pi\Sigma$ or  $\pi\Lambda$ channels below $\bar{K}N$ threshold. The
resulting $K^-p$ amplitude, extrapolated to subthreshold energies well below
the $\Lambda(1405)$, suggests strong attraction of the $K^-$ in a nuclear
environment. The following questions need to be addressed:\\ {\bf a)} Are such
binding effects sufficiently strong to compete with $\bar{K}N\rightarrow \pi
Y$ decay widths and $\bar{K}NN\rightarrow YN$ absorptive broadening ($Y =
\Lambda, \Sigma$)?\\ {\bf b)} Can $K^-NN$ or $K^-NNN$ clusters  be highly
compressed (as previously suggested by Akaishi and Yamazaki) in view of the
strong short range NN repulsion? 

We are investigating these questions using the AMD variational approach
(Akaishi, Dote et al.) for few-body systems, with a realistic Argonne V18
nucleon-nucleon potential and $\bar{K}N$ interactions derived from chiral
SU(3) dynamics. P-wave $K^-N$ interactions involving the $\Sigma(1385)$
resonance turn out to be important. For heavier nuclei, we solve the
Klein-Gordon equation with  $\bar{K}$-nuclear self-energies based on realistic
energy dependent input amplitudes.  
Our intermediate results are summarized as follows:     

{\bf 1.} $K^-pp$ {\it clusters}: quasi-molecular binding occurs despite the
strong short-range NN repulsion. However, the binding energy is extremely
sensitive to the poorly known range of the 
$\bar{K}N$ interaction. We note that the leading s-wave Weinberg-Tomozawa term
alone is not strong enough to produce binding which emerges only with
inclusion of $K^-p\leftrightarrow\pi\Sigma$ coupled-channels. The subthreshold
$\bar{K}N$ p-wave interaction supports binding significantly once the energy
drops below the $\Sigma(1385)$ resonance. $K^-pp$ binding energies in the
range around 100 MeV, together with widths larger than 60 MeV, are not
excluded. 

 {\bf 2.} $K^-ppn$ {\it and} $K^-pnn$ {\it systems}: weakly bound states
 appear to be possible, but their existence depends again sensitively on the
 range of the $\bar{K}N$ interaction. A previous interpretation (Akaishi and
 Yamazaki) of the observed KEK events in terms of narrow, deeply bound
 $K^-ppn$ and $K^-pnn$ states can so far not be confirmed within this extended
 framework. In particular, such states, if existent, are expected to have
 widths larger than 80 MeV. 
 
  {\bf 3.}  $K^-${\it nuclear bound states}: in nuclei from $^{16}$O through
  $^{208}$Pb the $K^-$ can be bound at normal nuclear density by about 50-80
  MeV, with widths of similar order. We find results in qualitative agreement
  with calculations by Mares, Friedman and Gal (reported at this workshop).

\newabstract %32 Yamazaki_deeplybound
\label{abs:Yamazaki_deeplybound}
 \begin{center} 
 {\large\bf Present status of the experimental investigation of deeply bound
 kaonic states}\\[0.5cm]  
 {\bf Toshimitsu Yamazaki}$^1$ and Yoshinori Akaishi$^2$\\[0.3cm] 
 $^1$Department of Physics, University of Tokyo, 7-3-1 Hongo, Bunkyo-ku, Tokyo
 113-0033, Japan, and Nishina Center for Accelerator-Based Science, RIKEN,
 Wako, Saitama 351-0198, Japan\\[0.1cm]  
 $^2$College of Science and Technology, Nihon University, Funabashi, Chiba
 274-8501, Japan, and Nishina Center for Accelerator-Based Science, RIKEN,
 Wako, Saitama 351-0198, Japan\\[0.1cm]  
 \end{center} 
  
We discuss the following problems with particular attention among others.

\bigskip
 
1) The $K^-$ capture at rest occurs in the nuclear surface region (not at a
   remote peripheral) by nucleons of substantial momenta, and thus, the
   emission spectra of proton and $\Lambda$ are broad, contrary to a recent
   claim of Oset and Toki \cite{OT}. Only one exceptional case is $K^-$
   capture by $^6$Li due to the very small internal momentum of $d$ in $^6$Li
   ($\sim$ 50 MeV/c). This explains the recently observed 500 MeV/c peak by
   FINUDA \cite{FINUDA-6Li} as originating from a quasi-$d$ capture, $K^- +
   ``d" \rightarrow p + \Sigma^-$, but such a monoenergetic proton peak is not
   expected from $^4$He nor from any other nucleus because the internal
   momentum of $d$ is large ($\sim$ 200 MeV/c). 

\bigskip
 
2) The invariant mass and the angular corrrelation of $\Lambda-p$ pairs
   emitted from $K^-$ capture by light nuclei as observed by FINUDA
   \cite{FINUDA-PRL} can be explained only by invoking a nuclear bound state
   $K^-pp$ \citetwo{Akaishi:02}{Yamazaki:02} with a binding energy of $B_K$ = 115
   MeV, but not by a suggested mechanism (via final state interactions) of
   Magas {\it et al.} \cite{MORT}. An improved experimental data in a wider
   range of the invariant mass is waited for.

\newabstract %33 Kishimoto
\label{abs:Kishimoto}
\begin{center}
{\large\bf Study of kaonic nuclei by in-flight $(K^-, N)$ reactions}\\[0.5cm]
Tadafumi Kishimoto \\[0.3cm]
Department of Physics, Osaka University
\\[0.1cm]
\end{center}

Study of kaonic nuclei becomes one of the central issues
recently since it could answer the question whether kaon
condensation takes place in the core of neutron stars.
We studied $\bar{K}$-nucleus system by the $(K^-, N)$ reaction
on $^{12}$C and $^{16}$O.  This reaction could be the best
probe to study $\bar{K}$-nuclear systems since reaction
mechanism is well understood.  The experiment was carried
out at the K2-beam line of 12GeV Proton Synchrotron at KEK.
$K^-$ beam of 1 GeV/c was employed for the study.  The observed
missing mass spectra were compared with theoretically calculated
spectra.  The comparison shows that the kaon nucleus potential
is as deep as 200 MeV.  We also observed large isospin
dependence which is consistent with I=0 dominance of KN
attractive interaction.  I described our experimental conditions
and current status of our analysis.

\newabstract %34 Herrmann
\label{abs:Herrmann}
\begin{center} 
 {\large\bf Search for deeply bound kaonic states with FOPI at GSI}\\[0.5cm] 
 N. Herrmann (FOPI collaboration)  \\[0.3cm] 
 Physikalisches Institut der Universi\"at Heidelberg, Philosophenweg 
 12, D-69120 Heidelberg, Germany 
\\[0.1cm] 
  \end{center} 
  
 The possibility that due to the strongly attractive kaon nucleon 
 potential deeply bound states might be formed 
 \cite{Akaishi:02}  
has triggered a number of experimental efforts and results 
 \cite{Suzuki:04} 
 that are  controversially interpreted \cite{OT05}.  
 More complete information 
 is needed from experiment. Based on the speculation that the 
 densities reached in heavy-ion reaction might be favorable for the 
 formation of such states\cite{Yamazaki:04}, a search program was initiated with 
 the FOPI apparatus at GSI \cite{RIT95b}.  
 
Exotic multi-baryon resonances might have a sizable 
decay branching ratio into the two body channel   $\Lambda$ - hyperons 
and protons (or deuterons) that can be easily reconstructed by a large 
solid angle charged particle detector like FOPI that allows to 
calculate the invariant mass of the particle pair.  The experimental 
difficulty arises from the fact that due to the very short lifetime of 
the resonance it will decay in the target and it's decay product are 
indistinguishable from directly emitted hadrons. With event samples of 
large statistics  correlated pairs can be identified by subtracting the 
background of uncorrelated ones. Data samples of 120 $\cdot 10^6$ 
events from the reaction  
Ni+Ni and 370  $\cdot 10^6$ Al + Al events at an incident energy of 
1.9 AGeV are available for the search. 
 
The background of 
uncorrelated pairs needs to be constructed very carefully in order to 
remove correlations in the data sample that originate from 
non-resonance effects like the presence of a reaction plane and 
tracking efficiencies of the detector. In order to demonstrate the 
feasibility of reconstructing short lived resonances, $\Lambda + \pi$ 
correlations are analyzed. The $\Sigma^*(1385)$ hyperon can be 
clearly identified with a significance of about 10 and a reconstructed 
width of $\Gamma = 50 \pm 10$\,MeV that is in agreement with the PDG value. 
The relative production yield with respect to the $\Lambda$ - baryon 
is 5 $\cdot 10^{-2}$ at a signal - to - background ratio of  3 $\cdot 
10^{-2}$. No stable signal like this could be identified so far in the  
correlations of $\Lambda$ - baryons with protons and deuterons.

\newabstract %35 Buehler
\label{abs:Buehler}
\begin{center}
 {\large\bf Search for K$^-$pp clusters in p+d-reaction with FOPI}\\[0.5cm]
 {\bf P.~B\"uhler}$^1$, M.~Cargnelli$^1$, L.~Fabbietti$^2$, P.~Kienle$^{1,2}$,
 N.~Herrmann$^3$, K.~Suzuki$^2$, T.~Yamazaki$^4$, J.~Zmeskal$^1$ \& FOPI
 collaboration \\[0.3cm]
 $^1$Stefan Meyer Institut, Vienna, Austria,
 $^2$Technical University M\"unchen, Germany,
 $^3$University Heidelberg, Germany,
 $^4$RIKEN \& University of Tokyo, Japan
 \end{center}

\begin{figure}[h]
 \begin{center}
 a)\includegraphics[width=55mm]{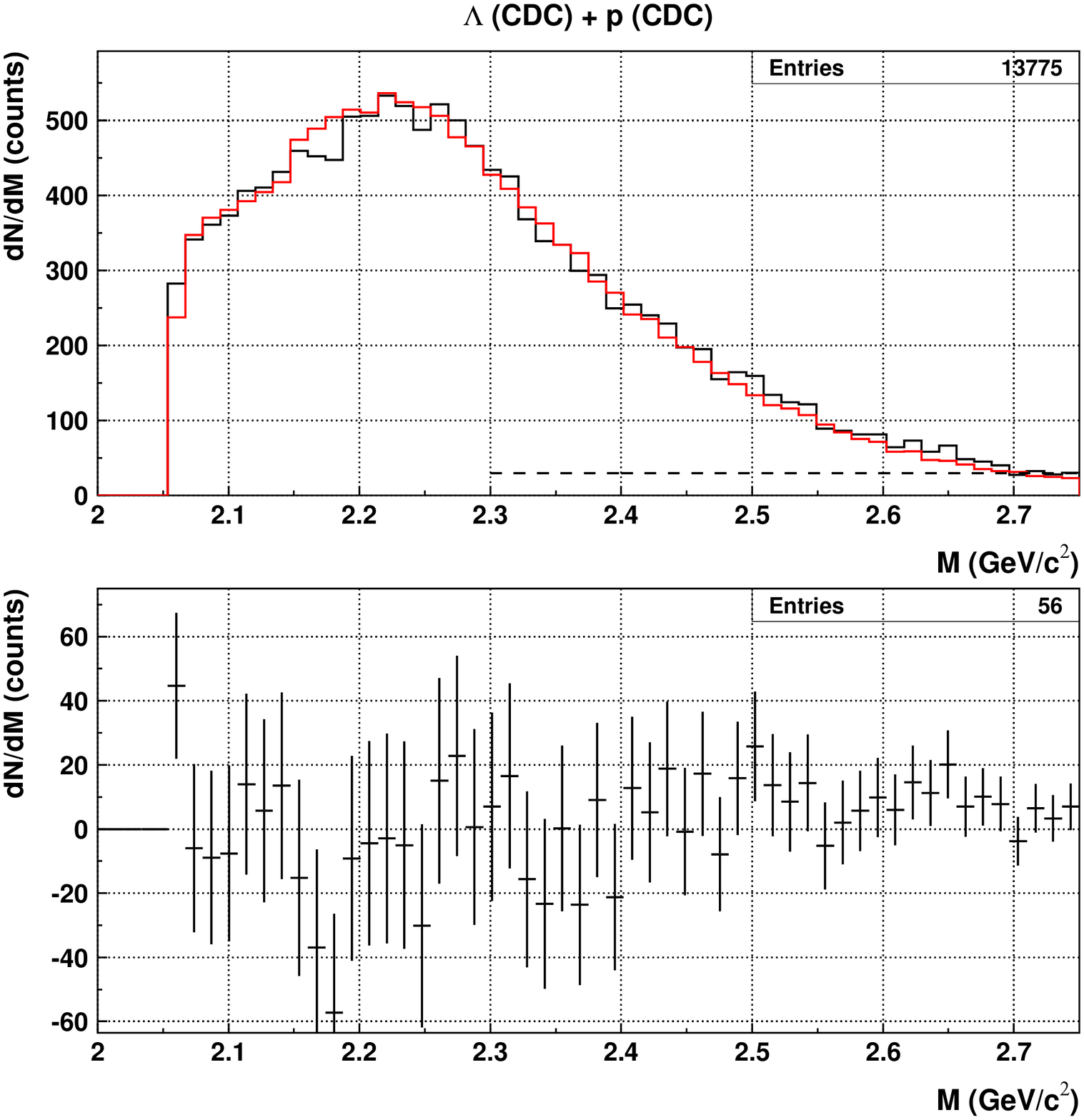}\hspace{20mm}%
 b)\includegraphics[width=55mm]{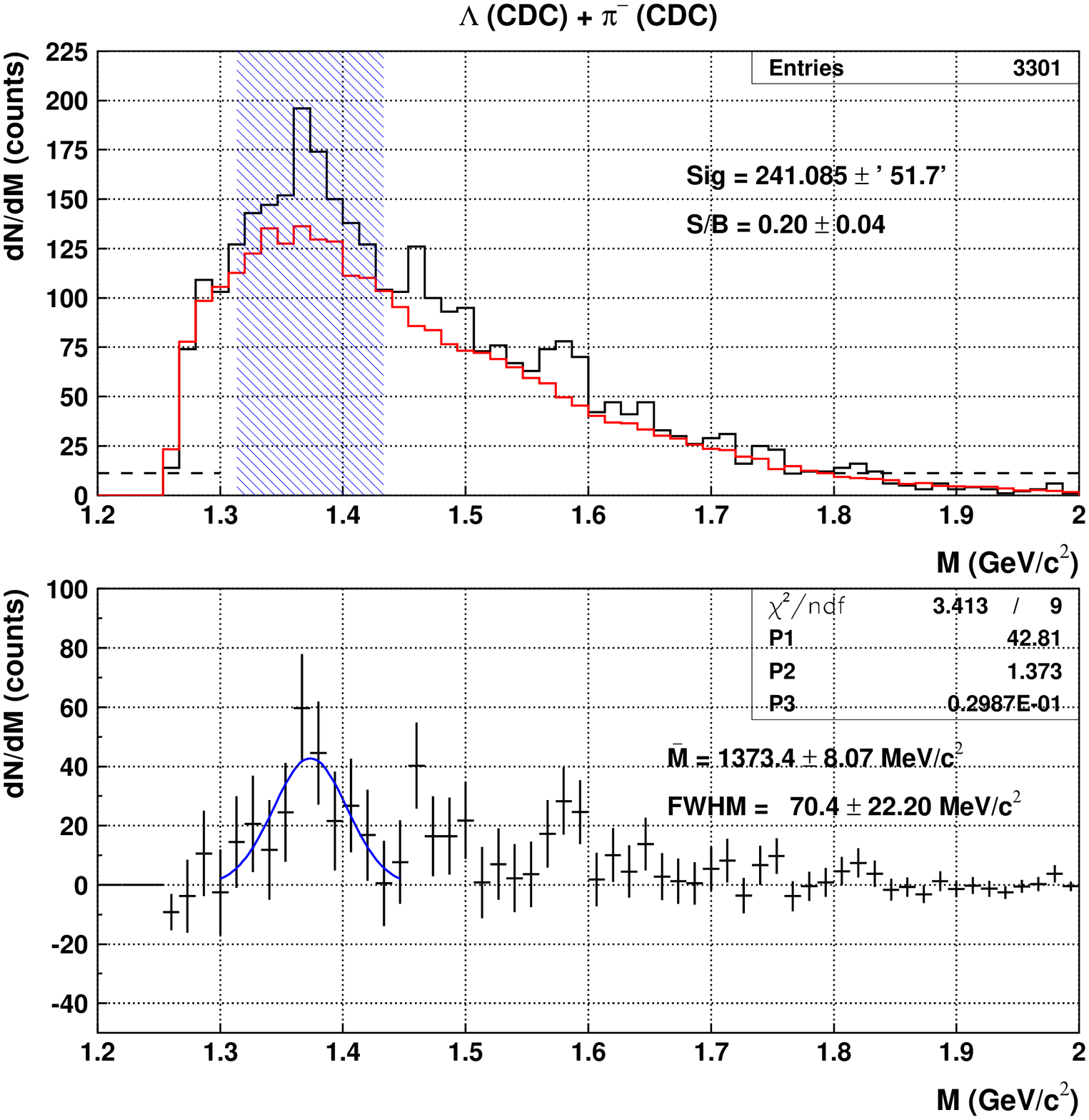}
 \end{center}
 \caption{Invariant mass of $\Lambda+p$ (left) and $\Lambda+\pi^-$. Upper panels
 show signal (black line) and combinatorial background (red line) and lower
 panels show background subtracted signal.}
 \label{fig01}
 \end{figure}

 In November 2005 an experiment was carried out with the FOPI detector at GSI
 with the aim to produce and verify the existence of the K$^-$pp, the simplest
 system of {\it "$\bar{K}$ nuclear clusters"}, in collisions of $3.5\;
 \mbox{GeV}$ protons with a liquid deuterium target. With the reaction $p + N
 \rightarrow p + \Lambda^* + K \rightarrow K^-pp + K \rightarrow \Lambda + p + K
 \rightarrow p + \pi^- + p + K$, proposed by Yamazaki \& Akaishi the existence
 of a $K^-pp$ cluster manifests as a peak in the invariant mass spectrum of the decay
 products of the $K^-pp$ ($\approx 2.28\;\mbox{GeV/c$^2$}$) and also in the
 missing mass spectrum of the reaction products. In combination these two mass
 spectra allow to unambiguously test the existence of these clusters. With the
 data gathered during this first run - limited statistics and data quality - it
 was however not possible to compute the missing mass spectra and only the
 invariant mass analysis was performed. Figure \ref{fig01}a) shows the invariant
 mass spectrum of $\Lambda + p$ measured in  the central drift chamber (CDC). In
 the upper panel the black line represents same-event combinations and the red
 line is the background estimated with an event-mixing method. The lower panel
 shows the difference and is consistent with no signal. With the available
 statistics only about $20$ K$^-$pp clusters are estimated to be formed, which
 is not detectable in the relatively large background. With the same technique
 also $\Lambda + \pi^-$ correlations can be studied (figure \ref{fig01}b)). In
 this case an enhancement is observed around the nominal mass of the $\Sigma$
 resonance at $1385\; \mbox{GeV/c$^2$}$. The exact peak position and width depend on
 the background treatment and need further systematic analysis. Bumps at masses
 of $1480$ and $1580\;\mbox{GeV/c$^2$}$ may be indications of other $\Sigma$
 resonances. A proposal for additional beamtime is in preparation. Experimental
 improvements are discussed which shall allow to suppress the combinatorial
 background. A more thorough investigation of the $\Sigma$-resonances with a
 higher-statistics run is envisaged as well.

\newabstract %40 Aslanyan
\label{abs:Aslanyan}
\begin{center}
 {\large\bf $\Lambda p$ spectrum analysis at 10 GeV/c in p+C interactions}\\[0.5cm]
 P.Zh. Aslanyan$^{1,2}$\\[0.3cm]
 $^1$Joint Institute for Nuclear Research, Dubna, Russia\\[0.1cm]
 $^2$ Yerevan State of University, Yerevan, Armenia\\[0.1cm]

 \end{center}

 On the theoretical side, many calculations of the $\Lambda p$
correlations  have been performed using  bag models\cite{jaffe}, a
phenomenological "Kaonic Nuclear Cluster models"   \cite{KNC}   and
et. al. New particles or states of matter containing 1,2-or more
strange  quarks have inspired a lot of experiments at  BNL(AGS),
CERN, FNAL, GSI, SEBAF, KEK, JINR and et al..

   The effective mass spectra of strange multiquark metastable states with $\Lambda$
   hyperon systems  from proton exposure in  pC $\to\Lambda$X reaction at 10
   GeV/c in 700000 
   stereo photographs (or neutron exposure at 7 GeV/c) on LHE JINR PBC
    were observed significant enhancement in invariant
   mass spectra\cite{exlp1}-\cite{exlp3}:($\Lambda$ p), ($\Lambda p \pi$),
   ($\Lambda$$\Lambda$),($\Lambda$,p,p) and   ($\Lambda \pi\pi$). There were
   succeeded in finding narrow resonance-like peaks  by 
   using different bin  sizes and conditions of the analysis for ($\Lambda$p)
   spectra 
     in ranges of :(2085-2120),(2145-2180), (2195-2230),(2260-2300)
     and   (2360-2400) MeV/c$^2$.
       A few events, detected on the photographs of the propane bubble chamber
      exposed to a 10 GeV/c proton beam, were interpreted as weak decays of H
   dibaryons \cite{exh1}-\cite{exh4}. 
          There are two groups of events interpreted as S=-2 stable dibaryons:
      1) the first group  is formed of  the  neutral, S=-2 stable dibaryons,
      the masses of which are below ($\Lambda,\Lambda$)   threshold;  2) the second group
       is formed  of neutral and positively charged S=-2 heavy stable dibaryons.
       The weak decay  mode of  dibaryon hypothesis  were observed by decay channels
       of $\Sigma^-p$,$\Lambda \pi^0p$, $\Lambda\pi^-p$, $\Sigma^+p\pi^-$ and  $K^-pp$.

\newabstract %41 Yamazaki_clusters
\label{abs:Yamazaki_clusters}
 \begin{center} 
 {\large\bf Enhanced formation of $K^-pp$ clusters by short-range $pp$
 collisions }\\[0.5cm]  
 {\bf Toshimitsu Yamazaki}$^1$ and Yoshinori Akaishi$^2$\\[0.3cm]  
 $^1$Department of Physics, University of Tokyo, 7-3-1 Hongo, Bunkyo-ku, Tokyo
 113-0033, Japan, and Nishina Center for Accelerator-Based Science, RIKEN,
 Wako, Saitama 351-0198, Japan\\[0.1cm]  
 $^2$College of Science and Technology, Nihon University, Funabashi, Chiba
 274-8501, Japan, and Nishina Center for Accelerator-Based Science, RIKEN,
 Wako, Saitama 351-0198, Japan\\[0.1cm]  
 \end{center} 
 
The most basic kaonic nuclear cluster, $K^- pp$, was predicted
 \citetwo{Akaishi:02}{Yamazaki:02}, and an experimental indication has been
 observed by FINUDA \cite{FINUDA-PRL}. Its formation by nucleon and heavy ion
 collisions is seached for by FOPI \cite{FOPI} at GSI.    
 We have found theoretically that the elementary process, $p + p \rightarrow K^+ + \Lambda(1405) + p$,  
which occurs in a short impact parameter and with a large momentum transfer
 ($Q \sim 1.6$ GeV/$c$), leads to unusually large self-trapping of
 $\Lambda(1405) (\equiv \Lambda^*)$ by the projectile proton, when a $K^-pp$
 system exists as a dense bound state. The seed, called ``$\Lambda^* p$
 doorway", is expected to play an important role in the ($p, K^{+,0})$ type
 reactions and heavy-ion collisions to produce various $\bar{K}$  nuclear
 clusters.

\begin{figure}[htb] 
\centering 
\includegraphics[width=7cm]{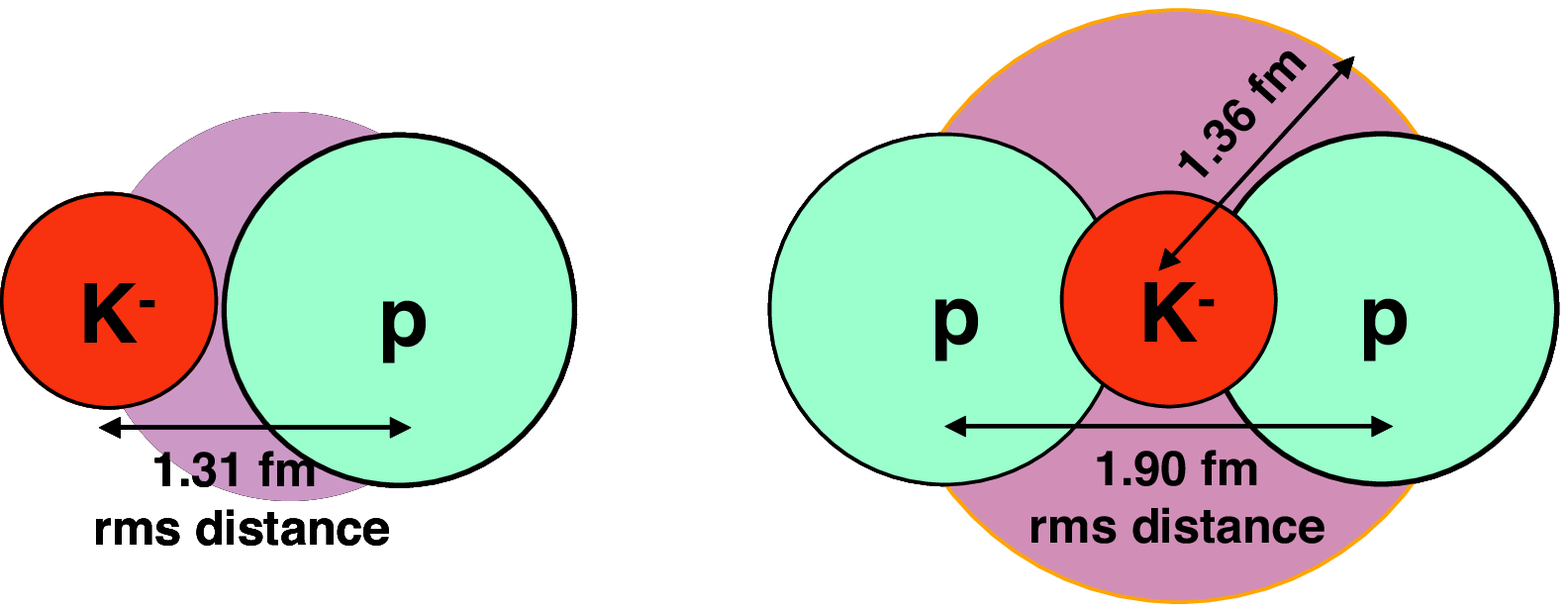} 
\includegraphics[width=7.5cm]{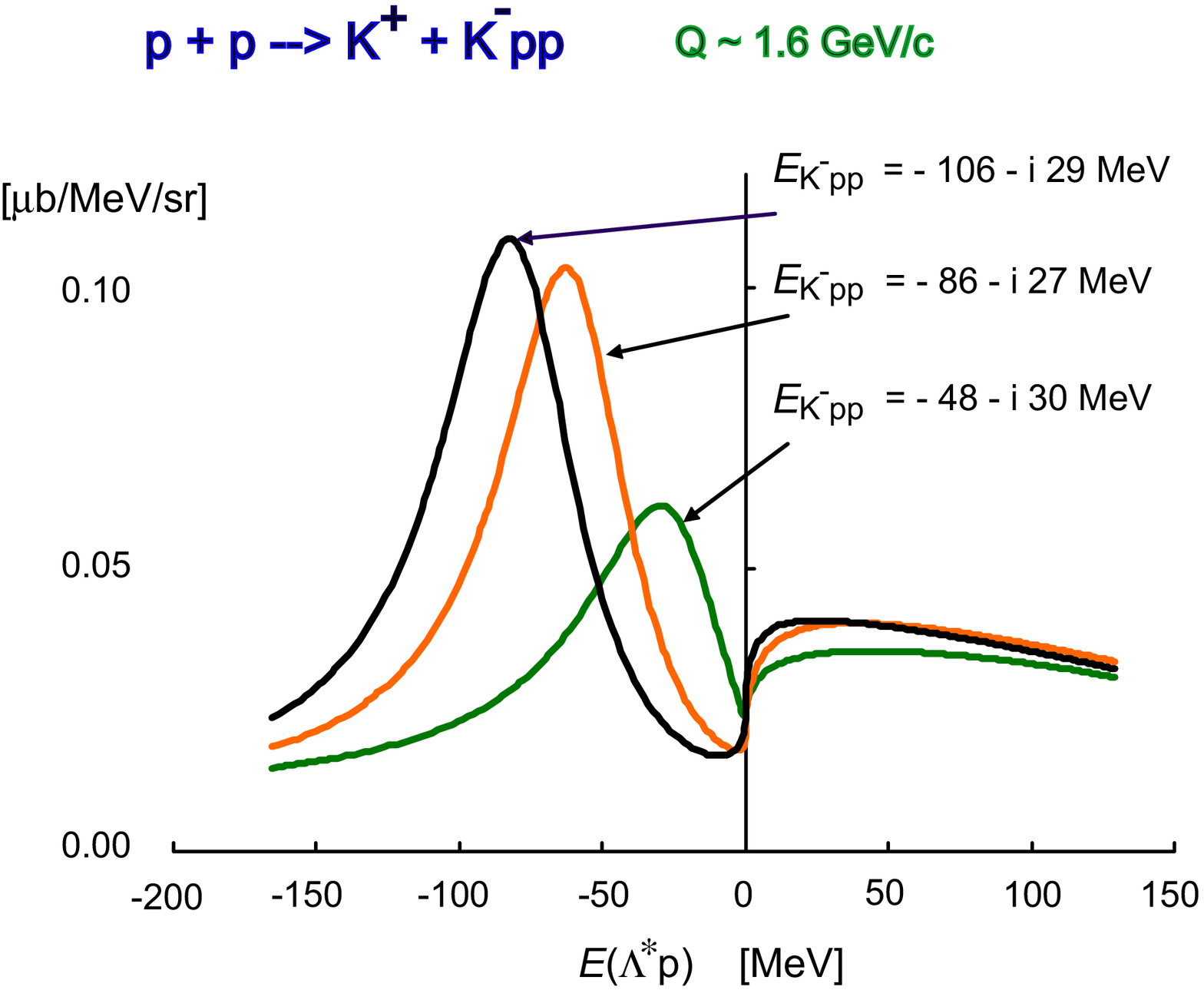} 
\vspace{0cm} 
\caption{\label{fig:Kp-Kpp}  
(Left) Predicted structure of $K^-p$ and $K^-pp$. (Right) Spectral shapes for
different binding energies of $K^-pp$ in $pp \rightarrow K^+ + K^-pp$ reaction
at $T_p$ = 3 GeV.  
} 
\end{figure}

\newabstract %42 Ohnishi
\label{abs:Ohnishi}
\begin{center}
{\large\bf A search for deeply-bound kaonic nuclear states 
by in-flight $^3$He(K$^-$,n) reaction at J-PARC }\\[0.5cm]
H. Ohnishi for J-PARC E-15 experiment \\[0.3cm]
 RIKEN, Nishina Center for Accelerater Based Science, 351-0198, Saitama,
Japan\\[0.1cm] 
\end{center}

We propose to perform an experimental search for deeply bound kaonic states
using $^3$He target by the in-flight kaon reaction, both the invariant mass
and missing mass spectroscopy with design resolution of 37 MeV/$c^2$ (FWHM)
and 20 MeV/$c^2$ (FWHM), respectively.  

Since, the interaction between $K^-$ and proton is confirmed to be strongly
attractive\cite{PRL78-3067}, one can assume that $\Lambda(1405)$ would be a
bound state between $K^-$ and proton. This assumption is naturally extended by
Akaishi and Yamazaki\cite{PRC65-044005} to the light nucleus, such as $^3$He,
$^4$He and $^8$Be, to investigate whether $K^-$ forms bound state with light
nuclei or not. Their coupled channel calculation predicted $K^-$ bound states
to have narrow widths and large binding energies.  

The KEK-PS E471\cite{PLB597-263} experiment was motivated with the possible
formation of deeply bound kaonic nucleus $K^-ppn$ with isospin zero, which is
expected to be detected most easily. The result is much different from the
prediction in mass (twice bigger in terms of binding energy) and in isospin
(T=1 instead of 0). To understand the difference between the data and the
theory, study on simple system would be most efficient. 

The second simplest kaonic nuclear system is $\bar{K}$ bound with two
nucleons, such as a $K^-pp$ state. Theoretically, binding energy and width is
calculated to be 48 MeV and 61 MeV\cite{PLB535-70},
respectively. Experimentally, this reaction allows to perform missing mass
study using primary neutron, and invariant mass spectroscopy via the decay
chain $K^-pp \rightarrow \Lambda p \rightarrow \pi^- pp$,
simultaneously. Detailed study of this simple system would be a doorway
towards investigation of the kaon bound states in heavy nucleus and/or a
multi-kaon bound system in a nucleus.

\newabstract %43 Zmeskal
\label{abs:Zmeskal}
\begin{center}
 {\large\bf AMADEUS  AT  DA$\Phi$NE }\\[0.5cm]
Johann Zmeskal on behalf of the AMADEUS Collaboration, \\[0.3cm]
% $^1$Laboratori Nazionali di Frascati dell' INFN, C.P. 13-I-00044, Frascati, 
% Italy\\[0.1cm]
% $^2$HISKP (Theorie), Univertit\"at Bonn, Nu\ss{}allee 14-16, 53115 Bonn, 
% Germany\\[0.1cm]
% $^3$
Stefan Meyer Institut f\"ur subatomare Physik, \"Osterreichische Akademie 
 der Wissenschaften, Boltzmangasse 3, A-1090, Wien, \"Osterreich
 \end{center}

A new series of experiments are planned at Laboratori Nazionali di
Frascati to search for the existence of antikaon{}-mediated bound
nuclear states [1] with an upgrade of the DA${\Phi}$NE machine [2].
This search deals with one of the most important, yet unsolved,
problems in hadron physics: how the hadron masses and hadron
interactions change in the nuclear medium and what might be the
structure of cold dense hadronic matter. Deeply bound antikaon nuclear
states ( ${\bar{{K}}}$ {}-nuclear clusters), if they exist,  will
offer the ideal conditions for investigating the way in which the
spontaneous and explicit chiral symmetry breaking pattern of
low{}-energy QCD changes in the nuclear environment [3]. 
 % example of embedding the figure below:

The design strategy for the search of antikaon{}-mediated bound nuclear
systems is to go fully exclusive -- that means not only to detect
the formation channel by missing mass spectroscopy, but also the decay
channel using invariant mass analysis. Therefore, an unambiguous
picture of a formed kaonic cluster could be extracted. 

To do so, it is necessary to build a detector which allows the
determination of all involved charged and neutral particles.
Fortunately, already a great part of the detector exists at
DA${\Phi}$NE, namely the KLOE detector with a large Central Drift
Chamber (CDC) for charge particle tracking, surrounded by an
Electromagnetic Calorimeter (EMC) optimized for gamma detection and
also providing neutron detection. The inner region around the beam pipe
has to be adapted for the AMADEUS setup with a cryogenic target and an
inner tracker system. 

The scientific program of AMADEUS consists of precision spectroscopy
studies, starting with light nuclei{}- $^3$He and $^4$He to form the 
most basic antikaon nuclear clusters: ``strange dibaryon'' 
($ppK^-$)  and ``tribaryon'' ($ppnK^-$, $pnnK^-$) states.
 Measurements of medium heavy nuclear targets are planned as well. A
detailed structure information can be extracted from a Dalitz analysis
of three{}-body decays of kaonic nuclei as was pointed out
recently by Kienle, Akaishi and Yamazaki [4]. This is one of the most
interesting feature to be performed with the AMADEUS setup. Finally,
these data will clearly proof or disproof the existence of
antikaon{}-mediated bound nuclear systems, a question strongly
discussed in theory.

\newabstract %44 Weise_summary
\label{abs:Weise_summary}

\begin{center}
 {\large\bf Discussion panel on deeply bound $\bar{K}$-nuclear states}\\[0.5cm]
 W. Weise (Convener)\\[0.3cm]
 Physik-Department, Technische Universit\"at M\"unchen, D-85747 Garching, Germany\\
 \end{center}
 
The panel session closing this workshop focused on the quest for deeply bound
states of antikaon-nuclear systems which, if existent, would open up an
entirely new dimension in the physics of hadrons and nuclei. The discussion
proceeded along a line of questions "Where do we stand?" and "What next?" 

\bigskip

{\bf Part I: Experiment}
 
The present experimental situation is not yet conclusive. Missing mass spectra
from $(K^-,p)$ and  $(K^-,n)$ with stopped kaons on $^4$He at KEK show signals
tentatively interpreted as $K^-pnn$ and $K^-ppn$ bound state candidates at
binding energies $B(K^-pnn) \simeq 194$ MeV and  $B(K^-ppn) \simeq 169$
MeV. Their small widths $(\Gamma < 20$ MeV) are, however, not understood. The
FINUDA experiment observes structures in $\Lambda p$ invariant mass spectra,
following stopped $K^-$ absorption on Li and C nuclei, which can
hypothetically be interpreted as $K^-pp$ clusters with $B(K^-pp) \simeq 115$
MeV and $\Gamma \simeq 70$ MeV. This interpretation competes with a
conventional analysis in terms of final state interactions. Searches are also
performed with FOPI at GSI for similar structures in spectra produced by ion
induced collisions with nuclei. Necessary next steps: 

- The KEK data require confirmation with higher statistics.

- Variation of kinematical cuts in the FINUDA experiment and systematics over
  several nuclei should clarify the role of final state interactions. 

- A detailed re-analysis of data taken with the KLOE drift chamber is performed in search for  
$K^-ppn$ and $K^-pnn$ clusters. 

- In the future the AMADEUS experiment at LNF is expected to provide a much
improved, high-statistics data base with special emphasis on exclusive
measurements of all final states.

\bigskip

{\bf Part II: Theory} 

The theoretical studies have so far been restricted to simple potential models
implemented in few-body variational calculations and mean-field
approaches. Improvements are required at several levels: 

- Realistic finite-range subthreshold $\bar{K}N$ interactions with their full
  energy dependence must be incorporated in the few-body computational
  framework. Accurate threshold constraints from kaonic hydrogen measurements
  (SIDDHARTA) will be important, beyond the precision that has already been
  reached with the DEAR experiment.  

- Realistic nucleon-nucleon interactions must be used with particular
  attention paid to short-range dynamics. 

- A detailed theoretical treatment of $\bar{K}NN \rightarrow YN$ absorption
  channels is essential in order to understand the widths of possible
  $K^-$nuclear strongly bound states. A high-precision determination of real
  and imaginary parts of the $K^-$deuteron scattering length (again a case for
  SIDDHARTA) will be important in determining the corresponding
  $\bar{K}NN-$to$-YN$ coupling. 

- In constructing a Hamiltonian for such systems, systematic guidance by an
  appropriate effective field theory is mandatory.

\bigskip

{\bf Part III: The Dense Matter connection}    

The existence of antikaon-nuclear bound systems may have an impact on
long-standing questions related to kaon condensation in dense
matter. Connections with the equations of state for nuclear matter (deduced
from high-energy heavy-ion collisions) and neutron star matter (from
progressively more accurate astrophysical observations) should be examined in
this context.

 \newabstract %36 Prades
\label{abs:Prades}
\begin{center} 
{\large\bf FSI in $K\to 3\pi$  
and Cabibbo's Proposal to Measure $a_0-a_2$}\\[0.5cm] 
Elvira G\'amiz$^1$, {\bf Joaquim Prades}$^2$,  
and Ignazio Scimemi$^3$\\[0.3cm] 
$^1$Department of Physics and Astronomy, The University of Glasgow, 
Glasgow G12 8QQ, United Kingdom\\[0.1cm] 
$^2$ CAFPE and Departamento de F\'{\i}sica Te\'orica 
y del Cosmos, Universidad de Granada, Campus de Fuente Nueva, E-18002 
Granada, Spain\\[0.1cm] 
$^3$ Departament de F\'{\i}sica Te\`orica, IFIC, CSIC-Universitat 
de Val\`encia, Apt. de Correus 22085, E-46071 Val\`encia, Spain 
\end{center} 
 
In this work \cite{GPS06}, 
 we study the recent Cabibbo's proposal to measure the $\pi\pi$  
scattering lengths combination $a_0-a_2$ from the cusp effect 
\cite{CAB04} in the $\pi^0\pi^0$ energy spectrum around threshold 
both for $K^+\to \pi^0\pi^0\pi^+$ and $K_L\to\pi^0\pi^0\pi^0$, 
and give the relevant formulas to describe it  including NLO in $\pi\pi$ 
scattering effects near threshold. 
We use fitted CHPT formulas at NLO to describe 
the real part of the regular contribution to 
the $K\to 3\pi$ amplitudes near threshold while the imaginary  and the 
non-regular parts are obtained just using unitarity and analyticity. 
  
Previous NLO results can be found in \cite{CI05} and  
 first experimental results have been also presented \cite{NA48}. 
 At present, the theoretical uncertainties 
dominate and it is interesting both to check them and study how  
to reduce them further. 
 Recently, a non-relativistic effective field theory 
in scattering lengths has been presented in \cite{CGKR06}. 
 
 For explicit formulas and our results see \cite{GPS06}. 
Here, we  just enumerate our main conclusions.  
1) If we make the same approximations done 
in \cite{CI05}, we agree analytically with the results there. 
2) The presence  and effects of the singularity at pseudo-threshold  
are discussed. 
3)  We estimate the theory uncertainty  
due to NNLO scattering effects to be around  5 \% 
and it is thus necessary to go to NNLO 
in $\pi\pi$ re-scattering effects to reduce/check this uncertainty. 
4) Several approximations done in \cite{CI05} are identified and  
quantified. Though they are individually negligible, they tend to go 
in the same direction and add up to around 2 \%. These 
approximations should be of course eliminated if one wants to reach the  
per cent level of uncertainty. 
5) At NLO, our final theoretical uncertainty is around 5 \%, if 
various small --from 1 \% to 2 \%-- 
 theoretical uncertainties are added quadratically.  
If these small theoretical 
 uncertainties are added linearly, one gets a theoretical uncertainty 
around  7 \%.

\newabstract %37 Collazuol
\label{abs:Collazuol}
\begin{center}
 {\large\bf Pion scattering lengths from the NA48/2 experiment at CERN}\\[0.5cm]
 G. Collazuol, on behalf of the NA48/2 collaboration\\[0.3cm]
 Scuola Normale Superiore and INFN Sezione di Pisa, Italy
 \end{center}

The NA48 experiment at CERN collected more than $3.1\times10^9$ $K^\pm\to\pi^\pm\pi^+\pi^-$
decays and more than $110\times10^6$ $K^\pm\to\pi^\pm\pi^o\pi^o$ decays for searching 
CP violating asymmetries in charged kaon decays. No asymmetries were found up to 
now~\citetwo{ASY_C}{ASY_N} at a level of $10^{-4}$ but an unexpected effect involving 
strong interactions was found while studing the kinematics of those decays.
A study~\cite{CUSP} of a partial sample of $23 \times 10^7$ $K^\pm\to\pi^\pm\pi^o\pi^o$ decays 
was reported at this workshop, whose results show an anomaly (cusp) in the
distribution of the $\pi^o\pi^o$ invariant  
mass ($m_{oo}$) in the region around $m_{oo} = 2 m_{+}$, where $m_+$ is the charged pion mass. 
This cusp, never observed in the past, is interpreted as an effect due mainly 
to the final state charge exchange scattering process $\pi^+\pi^-\to\pi^o\pi^o$ in the
$K^\pm\to\pi^\pm\pi^+\pi^-$ decay~\cite{CAB_1}. 

This provides a new method for a precise determination of the difference $a_0-a_2$ 
between the scattering lengths in the isospin $I=0$ and $I=2$ states. 
A best fit to a rescattering model~\cite{CAB_2} corrected for isospin symmetry 
breaking gives $(a_0-a_2) m_+ = 0.268\pm0.010_{stat}\pm0.004_{syst}$ 
with additional external error of $0.013$ from branching ratio and theoretical
uncertainties. The measured value is in agreement with chiral perturbation theory 
and with other measurements exploiting different methods.
For the first time the parameter $a_2=-0.041\pm0.022_{stat}\pm0.014_{syst}$ was 
directly measured.
If the correlation between $a_0$ and $a_2$ predicted by chiral symmetry is taken 
into account, this result becomes 
$(a_0-a_2) m_+ = 0.264\pm0.006_{stat}\pm0.004_{syst}\pm0.013_{ext}$.
Pionium bound state is also found as an additional sharp peak on top of the cusp.
By analyzing the full data sample we expect an increase in statistics by a factor of 5.  
An experimental error below $1.5\%$ seems not to be out of reach.  
At the moment the external uncertainty related to the theoretical method is $5\%$ 
and the data quality calls for additional theoretical effort in order to extract precise 
values of the $\pi\pi$ scattering parameters (higher orders and electromagnetic corrections).  
A fit according to different amplitudes representation~\cite{CGKR} is also in progress.

The cusp effect was confirmed by the NA48 collaboration, by finding a similar anomaly 
in the study of the $\pi^o\pi^o$ invariant mass in the $K_L\to3\pi^o$ decays, collected 
in the year 2000 with a statistics exceeding $100\times10^6$.

\newabstract %38 Isidori
\label{abs:Isidori}
\begin{center}
{\large\bf On the cusp effects in $K \to 3\pi$ decays}\\[0.5cm]
N. Cabibbo,${}^{a}$ {\bf G. Isidori}~${}^{b}$\\[0.3cm]
${}^{a}~$Dip.~Fisica and INFN, Univ.~Roma ``La Sapienza'', 
          P.le A.~Moro 2, I-00185 Rome, Italy \\
${}^{b}~$INFN - Laboratori Nazionali di Frascati, C.P.~13-I-00044, Frascati, 
Italy\\[0.2cm]
\end{center}

\noindent 
As pointed out in Ref.~\cite{Cabibbo:2004gq},
the rescattering of the final state 
pions produces a prominent cusp in the $M_{\pi^{0}\pi^{0}}$
spectrum of the $K^{+}\rightarrow\pi^{+}\pi^{0}\pi^{0}$ decay.
This effect can be 
used to obtain a precise determination of the 
$\pi$--$\pi$ scattering lengths ($a_I$) and, particularly, 
of the $a_{0}-a_{2}$ combination. 
In order to fully exploit the high-statistics and high-quality 
data collected by NA48 on the 
$K^{+}\rightarrow\pi^{+}\pi^{0}\pi^{0}$ decay~\cite{NA48},
an accurate theoretical description 
of this effect in terms of the $a_I$ 
is necessary. A first step in this direction has been 
presented in Ref.~\cite{CI}. 

The method of Ref.~\cite{CI}
is based on a systematic expansion 
in powers of the $\pi$--$\pi$ scattering lengths. 
The approach is less ambitious than the ordinary loop
expansion performed in effective field theories, 
such as CHPT: the scope is not a dynamical 
calculation of the entire decay amplitudes, 
but a systematical evaluation of the 
singular terms due to rescattering effects only. 
As far as the  description of the cusp effect
is concerned, this approach
is more efficient and substantially 
simpler than ordinary CHPT. 
Using this method, the coeff.~of the 
square-root singularities occurring at  
$M_{\pi^{0}\pi^{0}} = 2 M_{\pi^+}$
have been computed at $O(a^2_I)$ accuracy. 

\begin{figure}[h]
\vskip -2 mm
\begin{minipage}{55mm}
\includegraphics[width=5.2 cm]{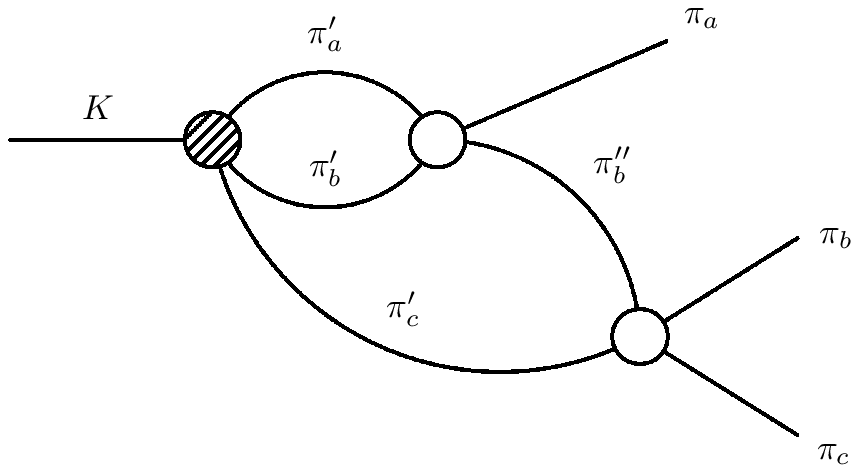} \hfill
\caption{\em 2-loop topology.\label{fig:2loop_CI}}
\end{minipage}
\begin{minipage}{110mm}
A technical assumption has been employed to simplify
  the calculation 
of complicated two-loop topologies of the type in Fig.~\ref{fig:2loop_CI}:
one-loop subdiagrams have been approximated by suitable 
polynomial expressions \cite{CI}. As recently shown in Ref.~\cite{BB},
this procedure does not reproduce the correct analytic 
properties of the amplitudes; however, it turns out to be an excellent
  numerical approximation. 
\end{minipage}
\vskip -3 mm
\end{figure}

\noindent
The theoretical error in the extraction of  $a_0-a_2$ from 
$K^{+}\rightarrow\pi^{+}\pi^{0}\pi^{0}$ based on the 
method of Ref.~\cite{CI} has been estimated 
to be $\approx 5\%$. The conservative nature of this estimate 
seems to be confirmed by the recent analysis of Ref.~\cite{GV}.
A somewhat larger uncertainty 
is expected in the $K_L \to 3\pi^0$ case, due to the accidental
smallness of the leading $O(a_I)$ term.
To go beyond this level of precision, a complete evaluation 
of radiative corrections is needed. The non-relativistic QFT 
formulation of Ref.~\cite{BB} offers a systematic 
tool to evaluate these effects.

\newabstract %39 Gasser
\label{abs:Gasser}
\begin{center}
 {\large\bf Non relativistic QFT and $K\rightarrow 3\pi$ decays}\\[0.5cm]
 {G.~Colangelo$^1$, {\bf J. Gasser}$^1$, B.~Kubis$^2$, 
 A.~Rusetsky$^{2,3}$}\\[.3cm]
$^1${Institute for Theoretical Physics, University of Bern,
Sidlerstr. 5, CH-3012 Bern, Switzerland}\\[.1cm]
$^2${Helmholtz-Institut f\"ur Strahlen- und Kernphysik,
Universit\"at Bonn, Nussallee~14-16, D-53115 Bonn, Germany}\\[.1cm]
$^3${On leave of absence from:
High Energy Physics Institute, Tbilisi State University,
University St.~9, 380086 Tbilisi, Georgia.}
\end{center}

Recently, it has been pointed out by 
Cabibbo and Isidori \citethree{cabibbo}{cabibboisidori}{isidori}
 that isospin violating effects
generate a pronounced cusp in $K\rightarrow 3\pi$ decays whose experimental
investigation may allow one to
 determine the combination $a_0-a_2$ of $\pi\pi$ scattering lengths 
with high precision. A
 first analysis of data based on this proposal 
has appeared \cite{NA48}.
 In order for this program to be carried through successfully,
 one needs to determine
 the structure of the cusp with a precision that matches the experimental 
accuracy. In view of the large amount of data available \cite{NA48}, this
is a considerable task. A first step in this direction has been done in
Ref. \cite{cabibboisidori}. For a confirmation of the 
uncertainties quoted in \cite{cabibboisidori}, 
see \citetwo{gamizetal}{prades}. In Ref. \cite{nrqft}, a non-relativistic 
QFT framework is
constructed, which automatically satisfies unitarity and analyticity 
constraints and, in
addition, allows one to include electromagnetic contributions in a standard
manner.
In this framework -- in contrast to relativistic field theory -- an
expansion in powers of scattering lengths emerges automatically from the
loop expansion. Moreover, it is a scheme that provides a proper power
counting.

The results in \cite{nrqft} are displayed without a detailed derivation, 
which will be provided in a forthcoming publication \cite{long}. A short
comparison with the framework  used in Ref.~\cite{cabibboisidori} 
is provided by Isidori in his contribution
to this workshop \cite{isidori}.

\end{document}